\RequirePackage{fix-cm}
\documentclass[twocolumn]{svjour3}          % twocolumn
\smartqed  % flush right qed marks, e.g. at end of proof
\usepackage{graphicx}

\usepackage{breqn} % for multi line equations
\usepackage{supertabular} % for our list of symbols which goes over two columns
\usepackage{natbib}
\usepackage[colorlinks=true,allcolors=blue]{hyperref}
\usepackage{epstopdf, epsfig}
\usepackage[]{subcaption} % options for left-aligned subcaption letters: singlelinecheck=off,justification=raggedright
\journalname{}
\begin{document}

\title{Sinusoidal-gust generation with a pitching and plunging airfoil}
%\subtitle{Controlling turbulence length scales, Reynolds number, isotropy, 
%and decay rate using active grid correlation kernels.  }


\author{Nathaniel J. Wei$^{1,2}$ \and Johannes Kissing$^1$ \and Cameron Tropea$^1$$^{\dagger}$\thanks{\noindent$^{\dagger}$Corresponding author. \email{tropea@sla.tu-darmstadt.de}}}

%\author{Author XXX\textsuperscript{a}\thanks{\noindent\textsuperscript{a}Address of the author XXX \\ Email...}\, Author YYY \textsuperscript{b}{\footnote{\textsuperscript{b} Address of the author YYY}} }

%Kevin\footnotemark% and Nathan\footnotemark[1]
%\footnotetext[1]{Hi testing 123}
%\renewcommand{\thefootnote}	{\fnsymbol{footnote}}
%\footnote[4]{text. This is a footnote.}
\authorrunning{N. J. Wei, J. Kissing, \& C. Tropea} % if too long for running head

\institute{$^1$Institute for Fluid Mechanics and Aerodynamics\\
Technische Universit\"at Darmstadt\\
Flughafenstrasse 19\\
64347 Griesheim, Germany\\
$^2$Department of Mechanical Engineering\\
Stanford University\\
Stanford, California 94305, USA
}

\date{Received: date / Accepted: date}

\maketitle

%===============================================%
%============= ABSTRACT ========================%
\begin{abstract}
% \label{abstract}
%===============================================%
%Should be 150-250 words. should not contain any undefined abbreviations or unspecified references.

\qquad The generation of uniform, periodic gust disturbances in an experimental context is demonstrated using a single oscillating airfoil. A pitching and heaving symmetric airfoil is suggested as a simpler alternative to existing gust-generation methods. The Theodorsen theory of unsteady aerodynamics is used as an analytical tool to dictate the kinematics necessary to produce well-defined sinusoidal gusts downstream of the airfoil. These analytic predictions improve the symmetry of fluctuations in the vertical velocity induced by the airfoil, as well as minimize the influence of vorticity shed by the oscillating airfoil. The apparatus is shown to produce smooth, repeatable gusts with high amplitudes and reduced frequencies compared to other gust-generation mechanisms in the literature. Furthermore, the control of downstream flow properties by airfoil motion kinematics has applications in experimental aerodynamics, the design of rotorcraft and light aerial vehicles, and biological propulsion.

%Include keywords, PACS and mathematical subject classification numbers as needed.  
% \PACS{47.27.Rc turbulence control \and 47.27.wb turbulent wakes}
% \subclass{76F70 control of turbulent flows}
\end{abstract}

%===============================================%

\section{Introduction}

The generation of well-defined periodic flow disturbances in a laboratory environment is a critical requirement for experiments involving unsteady aerodynamics, aircraft control, gust disturbances, unsteady sensor calibration, and many other applications. Modeling a periodic disturbance in simulations is comparitively easy: one must only specify a fluctuating velocity as an inflow condition, and the effects of the resulting flow can be studied directly. In experiments, however, generating the desired inflow conditions can be significantly more difficult. For example, gust disturbances of uniform character can be created from the turbulent fluctuations shed behind a passive grid \cite[e.g.][]{lysak_measurement_2016}. Also active grids have demonstrated the ability to control the intensity and structure of turbulence in experimental settings \cite[cf.][etc.]{makita_realization_1991, cekli_tailoring_2010, knebel_atmospheric_2011, griffin_control_2019}. These methods generate gust disturbances over a continuous spectrum. In many cases however, it is desirable to produce disturbances at a single frequency. It is relatively easy for example, to generate vortical structures at a given frequency. Vortex shedding behind large cylinders or pylons yields periodic fluctuations in velocity that can be used in experiments \cite[cf.][]{larose_experimental_1999}. The vortices shed by an impulsively plunging plate can also be used to represent gust disturbances \citep{hufstedler_vortical_2019}.

For several applications a sinusoidal gust is required. The Sears function, for example, models  unsteady loads on an airfoil for a sinusoidal gust disturbance in the normal velocity component \citep{sears_aspects_1941}. \cite{goldstein_complete_1976} and \cite{atassi_sears_1984} extended this theory to include sinusoidal velocity fluctuations in the streamwise as well as normal directions. Sinusoidal gusts are also useful for tests of aircraft dynamics and controls, in which a series of well-defined, single-frequency disturbance inputs allows the frequency response of the system to be established \citep{bennett_wind-tunnel_1966}. The calibration of pressure transducers for unsteady conditions may also be achieved using  a smooth, sinusoidal velocity signal. 

The usefulness of sinusoidal gust disturbances in experimental investigations has led to the development of several methods of generating them. One of the earliest and simplest generators, constructed by \cite{hakkinen_theoretical_1957}, involved a plunging plate in a wind tunnel. The apparatus was used in an attempt to experimentally validate the Sears function, but proved unsatisfactory due to sensor noise. Subsequent gust-generation systems, possibly inferring that a single actuated plate was not sufficient for producing well-defined gusts, became more and more complex. \cite{bennett_wind-tunnel_1966} used four plates, mounted in pairs on the walls of a wind tunnel and actuated together by a series of linkages, to produce sinusoidal gusts for experiments with scale models of aircraft. \cite{ham_wind_1974} and \cite{jancauskas_aerodynamic_1986} generated gusts using a pair of controlled-circulation airfoils. The concept was extended by \cite{tang_experiments_1996} in an array of four such airfoils. Approaches using arrays of six or more vanes \cite[e.g.][]{saddington_characterisation_2015, cordes_note_2017} have been tested, though the wakes of the vanes introduced turbulent fluctuations into the downstream flow conditions. Simpler generation mechanisms involving two pitching plates \cite[e.g.][]{lancelot_design_2015, wood_new_2017} avoid wake effects by construction, but tend to be limited in both the amplitude of the gusts produced (usually less than $1^\circ$) and reduced frequency they can achieve. However, this design has been proven to be effective in transonic wind tunnels \citep{brion_generation_2015}. Alternatively, a single pitching airfoil has been demonstrated in many contexts to be sufficient for generating vortical gust disturbances, where smooth, sinusoidal disturbances are not required and wake effects do not need to be controlled \citep{klein_high-lift_2014, klein_two-element_2017}.

The question of generating well-defined gusts using a single oscillating plate or foil should not be resolved by mere convention or consensus, especially since there are several advantages to such an apparatus for experiments. A single foil produces far less blockage in a wind tunnel than more complicated gust-generation devices. The actuation is also far simpler and less costly, and facilitates operation, maintenance, and removal. Lastly, the single actuation mechanism of the foil means that, in theory, the structure of the generated gusts can be more precisely controlled.

In this paper,  a theory for the generation of sinusoidal gusts with a single oscillating foil is proposed and validated  with wind-tunnel experiments. The theory is developed from analytical, physical, and experimental arguments. Its effectiveness in producing tailored periodic profiles that exhibit minimal disturbance from the wake of the foil itself is then demonstrated. This gust generator is able to produce sinusoidal gusts over a range of frequencies and amplitudes and lends itself  to experiments that require well-defined unsteady conditions.

\section{Theoretical Considerations}

In this section, several considerations from the literature regarding the wakes of oscillating airfoils are discussed. These ideas undergird the hypothesis that the generation of well-defined periodic gusts is possible with a single airfoil as a gust generator. A theory for the generation of these gusts is then developed from the Theodorsen model for unsteady aerodynamics. These theoretical considerations provide guidance for the experiments that  follow.

\subsection{Considerations from the literature}
\label{sec:sec2_literature}

In order to generate sinusoidal gusts with a single airfoil,  inspiration is derived from studies regarding the aerodynamics of oscillating airfoils. \cite{anderson_oscillating_1998} characterized the wake patterns behind a NACA-0012 airfoil actuated in pitch and plunge over a range of reduced frequencies, $k = \frac{\omega c}{2U_\infty}$, and Strouhal numbers, $St = \frac{fA}{U_\infty}$. Here $\omega = 2\pi f$ is the frequency of oscillation in radians per second, and $A$ represents the characteristic width of the wake, estimated by the plunge amplitude or the trailing-edge amplitude of the airfoil. For a range of combinations of pitch and plunge amplitudes, several distinct wake regimes were identified. Of particular interest to this study is the region defined by Strouhal numbers less than $0.2$ and pitch amplitudes less than $50^\circ$, in which a ``wavy wake" without significant vortex shedding was observed. This result suggests that the formation of well-defined sinusoidal oscillations in the wake of an airfoil can be achieved systematically as a function of pitch and plunge kinematics.

Two subsequent studies analyzed the effects of airfoil kinematics on wake structures generated by a pitching and plunging airfoil. For pitching and plunging airfoils, these wake structures are predominantly formed in dynamic stall, where strong leading-edge vortices (LEVs) and trailing-edge vortices (TEVs) roll up on and behind the airfoil, and convect downstream into the wake \citep{carr_progress_1988}. \cite{rival_influence_2009} studied the effects of non-sinusoidal actuation profiles on the propagation of these structures into the wake, demonstrating that the strength of the LEV on the airfoil and its persistence into the wake were affected by the shape of the motion waveforms followed by the airfoil. Extending this result, \cite{prangemeier_manipulation_2010} demonstrated that the addition of a quick pitch-down motion in the downstroke of an oscillating airfoil served to decrease the circulation of the TEV by as much as 60\%. The additional pitch motion of the airfoil interfered with the formation of the TEV, so that the strength of the TEV was not significantly dependent on the characteristics of the LEV. Similar reductions in wake vorticity were demonstrated in numerical simulations by \cite{gharali_dynamic_2013}. Therefore, though the strengths of LEVs and TEVs are inherently connected, the addition of pitching motions for a plunging airfoil can inhibit the generation of large trailing-edge vorticity in spite of large accumulations of vorticity at the leading edge. This is instructive to the current study, because the shedding of large, intermittent velocity fluctuations is antithetical to the generation of smooth, periodic velocity signals downstream.

In order to minimize vorticity effects in the flow downstream of the gust generator, the wake of the generator itself should be minimized along with the persistence of dynamic effects. To this end, \cite{hufstedler_vortical_2019} developed a vortical gust generator using an impulsively plunging plate. The plunging motions of the plate meant that the wake vorticity from the plate  had a limited effect on the gust region, since the wake vorticity would only cross through the measurement domain twice per cycle. This is in contrast to gust generators that employ purely pitching vanes, which are fixed in place and thus continually disturb the downstream flow conditions.

These considerations, taken together, imply that the generation of well-defined sinusoidal gusts is possible with a single airfoil, provided the kinematics of that airfoil are carefully controlled. Relatively low Strouhal numbers and reduced frequencies are required to avoid excessive vortex shedding. Plunge oscillations can help reduce the effects of the airfoil's wake on the downstream flow properties. Adding pitch motions can further attenuate trailing-edge vorticity, thereby reducing intermittency in the downstream velocity signals. The remainder of this work is therefore concerned with the combination of these observations: specifically, which combinations of pitch and plunge produce satisfactorily smooth sinusoidal velocity fluctuations for a range of Strouhal numbers and reduced frequencies. In order to achieve these gust conditions, intermittent effects from vortices shed by the gust generator should be avoided, and the resulting velocity profiles should be as symmetric and regular as possible. These criteria will drive the development and characterization of a robust theory for sinusoidal-gust generation with a single airfoil.

In this study, gust fluctuations in the flow-normal direction are specifically investigated. Fluctuations in the streamwise velocity and in the flow vorticity were not central to the applications of this work, and measurements of these quantities demonstrated that they were insensitive across the range of parameters tested in these experiments. Additionally, \cite{wei_insights_2018} have shown that, in the case of the Sears function, streamwise velocity fluctuations are not critical to obtaining experimental convergence with analytical predictions. Thus, gust fluctuations in the flow-normal direction are the main quantity of interest in this work.
% \todo{Are we allowed to cite this?, Maybe the problem will solve itself}

\subsection{Adaptation of the Theodorsen theory for the generation of sinusoidal gusts}
\label{sec:sec2_theodorsen}

\cite{theodorsen_general_1934} developed a first-order theory for the forces and moments experienced by a pitching and plunging flat-plate airfoil. The theory combines added-mass forces from the pitch ($\theta$) and plunge ($h$) motions of the airfoil with the influence of the circulation in the wake to predict the unsteady aerodynamic loads on the airfoil. Since the theory accounts for the effects of both kinematics and wake circulation, it can be used to build a relation for the kinematic parameters needed to generate smooth, periodic gusts. It has also been validated in the same experimental apparatus used in the present study \citep{cordes_note_2017}. The theory was used by \cite{brion_generation_2015} to estimate the amplitude of gusts produced by a purely pitching airfoil in transonic flow. Apart from this, it has been absent from the gust-generation literature. 

The pitching moment on the airfoil is of particular interest because it provides a connection between airfoil dynamics and downstream flow characteristics. While the Theodorsen theory does not model dynamic effects, the vortices associated with dynamic stall should not be present in the ideal gust-generation case, and thus the forces and moments on the airfoil should be captured well by the Theodorsen function. In this ideal case, the airfoil would simply redirect flow up and down, creating vertical fluctuations in velocity without introducing significant circulation into the downstream region. The leading edge of the airfoil would minimally disturb the incoming flow, so that the flow on the upper and lower surfaces of the airfoil follow the airfoil's surface without significant differences between the upper and lower surface. The resulting lack of velocity gradients at the trailing edge of the airfoil would then minimize the circulation injected into the wake. In this sense, both the added-mass moments and moments due to wake circulation would be optimally small, and the total pitching moment on the airfoil would be minimized.

The kinematics required to achieve this ideal gust-generation case can be calculated using Theodorsen's theory. For a given plunge amplitude $h_0$, inflow velocity $U_\infty$, and gust frequency $f$, it is possible to find the pitch amplitude $\theta^*$ that will minimize the amplitude of the pitching moment on the airfoil. The plunge amplitude is defined in this case by the amplitude of the trailing edge of the airfoil, as in \cite{anderson_oscillating_1998}, so that the length scale within the Strouhal number approximately represents the width of the airfoil's wake. The pitching moment is also defined about the trailing edge, in order to quantify the cumulative influence of the airfoil on the flow from the leading edge up to the trailing edge. Sinusoidal profiles for $h$ and $\theta$, along with their first and second derivatives in time are assumed, as the generated gust is desired to be sinusoidal as well. This collection of parameters is sufficient to compute the value of $\theta$ at which the pitching moment about the trailing edge is minimized, and, by extension, gusts of optimally sinusoidal character are produced.

However, a minimum pitching moment does not exist for positive $\theta$. This implies that the generation of sinusoidal gusts requires the pitch waveform to be precisely out of phase with the plunge waveform. When this correction is made, the pitching moment can be minimized for any given set of input parameters. Optimal pitch amplitudes for several sets of airfoil kinematics are given in Tables \ref{tab:pitchAmpsOptimizedCases} and \ref{tab:pitchAmpsBaselineCases} in Sec.\ \ref{sec:sec4_gustgen}.

This series of assumptions and arguments  for gust generation are conceptual by nature and require experimental investigation to verify whether they apply in practice. The character of the gusts produced by such a gust generator provides the validation criteria for the theory. In particular, the sinusoidal character of the velocity profiles and the extent of the wake of the airfoil downstream of the gust generator will serve as indicators for the quality of the gusts generated according to these principles.

\section{Experimental Setup}

In this section, the experimental apparatus used in this study is described in detail. The processing of raw data from the experiment is also outlined, and two metrics for quantifying the character of gusts are explained.

% \todo[inline]{Should we talk about how we were operating on the edge of what was possible with our equipment for these experiments? That would either justify the noise in the data (good), or cast doubt on our conclusions (bad). JK: Due to the fact that we are already using one of the most powerful PIV lasers available, mentioning the laser power to be the limiting factor for time-resolved quantification within such a large FOV is reasonable from my point of view. We could mention it at the mean acceleration and wake with discussion and point out, that we are investigating a fluctuating quantity obtained by spatial / temporal derivation of vector fields. We could also mention in the discussion of the respective figures, where the signal to noise ratio might have been too low. E.g. figure 11 (a)}

\subsection{Experimental apparatus}
\label{sec:sec3_apparatus}

\begin{figure*}[!ht]
    \includegraphics[width=\textwidth]{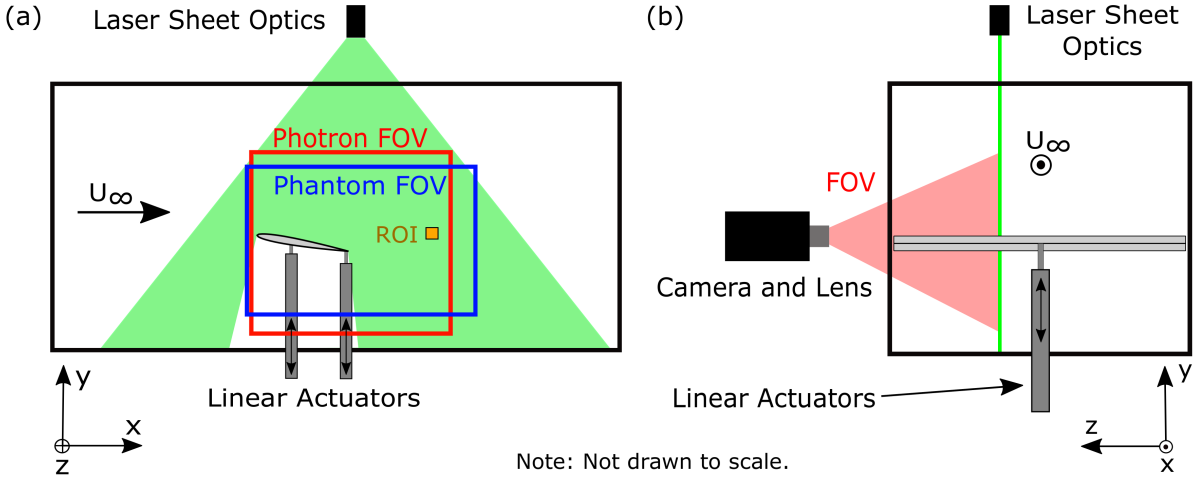}
    \caption{Schematic of the experimental setup: (a) side view, and (b) rear view. Fields of view of the two cameras are shown in (a). The region of interest (ROI) used to measure velocity and vorticity profiles is shown as an orange square in (a).}
    \label{fig:expsetup}
\end{figure*}

The experiments described in this work were carried out in an open-return wind tunnel at the Technische Universit{\"a}t Darmstadt. This tunnel had a square test section with a side length of 0.45 m. Optical access was provided by a removable glass plate that spanned the length of the test section, as well as a window on the top of the test section through which the laser sheet for the PIV system was passed. The flow velocity in the tunnel was held at $U_\infty = 2.5 \pm 0.1$ m/s using a closed-loop controller and an impeller anemometer (type TS16/15GE-
mc40A/125/p0/ZG1) from H{\"o}ntzsch GmbH installed near the inlet of the test section. This corresponded to a chord-based Reynolds number of about 20,000. Turbulence intensities in the test section were measured to be less than 2\%. More details on the wind tunnel can be found in \cite{rival_influence_2009}.

The test airfoil used in this study was a symmetric NACA-0008 profile with a chord of $c = 120$ mm. The airfoil was 3D-printed and reinforced in the spanwise direction with two carbon-fiber struts to inhibit elastic deformations. Two halves were printed, laminated in black foil to reduce reflections from the PIV laser light sheet, and attached to an aluminum mount, which was also laminated in black foil. The mount was supported by two LinMot PS01-48x240F-C linear actuators, having a dynamic position accuracy of $\pm 0.1$ mm. These were controlled from a computer and could be used to actuate the airfoil in pitch and plunge at frequencies of up to 10 Hz. The downstream actuator was centered on the airfoil's trailing edge. A schematic of the actuators and test airfoil is given as Fig.\ \ref{fig:expsetup}.

Flow-field measurements in this study were acquired using particle-image velocimetry (PIV). Seeding particles were generated from Di-Ethyl-Hexyl-Sebacat (DEHS), which had a mean diameter of 0.5-1.5 $\mu$m and a response time of $\tau_s = 2.7$ $\mu$s \citep{raffel_particle_2007}. The laser sheet was produced by a Litron LDY-303 high-speed dual-cavity Nd:YLF laser with a wavelength of 527 nm and frequency of 1 kHz. The sheet was generated by cylindrical lenses mounted high above the test section, in order to provide as large a field of view as possible. It was positioned at the edge of the optical-access window at the top of the test section, corresponding to a location of 20\% of the airfoil's span from the midpoint. Due to the power of the laser, an additional strip of fluorescent foil was wrapped around the airfoil at the location of the laser sheet, to further reduce reflections in the images and to prevent the laser sheet from burning into the airfoil itself. The location of the laser sheet is shown in Fig.\ \ref{fig:expsetup}. A larger field of view was not possible, as the laser had to be operated at full power in the given configuration in order to sufficiently illuminate seeding particles in the flow.

Images were captured at a frequency of 1000 pairs per second using a high-speed camera. In the first round of experiments, a Phantom v 12.1 camera was used. Repairs on the lab's Photron SA 1.1 camera were complete for the second round of experiments. A band-pass filter was applied with the Photron camera to further protect the camera sensor from reflections. For both cameras, a Zeiss 50-mm lens was used. The approximate  fields of view for both cameras are shown in Fig.\ \ref{fig:expsetup}(b). The cameras were mounted on a traverse, and the scale was computed using a target placed in the laser sheet. The camera, laser cavities, and actuators were all controlled with a LabVIEW script using a National Instruments PCI-6220 data-acquisition card. This allowed the measurements and data collection to be centrally synchronized. Additional details regarding differences in the setup between the two rounds of experiments are given in Table \ref{tab:params}.

\subsection{Data processing and analysis}
\label{sec:sec3_processing}

Two-component velocity fields were obtained from the raw images using the PIVview2C software package from PIVTEC GmbH. In both cases, a multigrid approach with three iterations was applied, with 50\% overlap between correlation windows. Window sizes were selected so that, on average, about 10 particles could be found in each interrogation area \citep{tropea_springer_2007}. Outlier detection and median filtering with $3 \times 3$ px windows were executed on the resulting vector fields. Specific details regarding processing parameters for the two rounds of experiments are given in Table \ref{tab:params}.

\begin{table}[!ht]
\centering
\begin{tabular}{| p{2.5cm} | p{2cm} | p{2cm} |}
\hline
Property & Round 1 & Round 2 \\ \hline\hline	
Camera & Phantom v12.1 & Photron SA1.1 \\ \hline
Resolution (px) & $1280 \times 800$ & $1024 \times 1024$ \\ \hline
Scale (px/mm) & 3.630 & 2.833 \\ \hline
Frame-straddling interval ($\mu$s) & 400 & 450 \\ \hline
Cross-correlation window size (px) & $32 \times 32$ & $16 \times 16$ \\ \hline
Peak-finding algorithm & $3 \times 3$ Gaussian fit & Whittaker reconstruction \\ \hline
\end{tabular}
\caption{Camera and PIV processing parameters for the two rounds of experiments presented in this work.}
\label{tab:params}
\end{table}

The location of interest within the flow field for these experiments was a point located one full chord length downstream of the trailing edge of the airfoil. This was selected with an eye toward future experiments: the experimental setup had been originally designed to study two oscillating airfoils in a tandem configuration \citep{rival_vortex_2011}, and the location of the leading edge of the downstream airfoil was 120 mm downstream of the trailing edge of the forward airfoil. Additionally, one chord length into the wake represented an intermediate distance at which the influences of neither transient effects nor viscous dissipation were significant. In future studies, it may be useful to investigate whether the results presented here hold in the far-wake region. For all of the results in this work, a box of length 15 mm was isolated at $x = 1.0c$ behind the test airfoil for the purpose of characterization of the gust signal, through averaging all velocity vectors within the region. This dimension was found to produce a stable average while still maintaining relatively high spatial resolution. This region of interest (ROI) was located at the same height as the midpoint of the plunge amplitude of the airfoil, defined as $y = 0$, and is represented schematically as an orange box in Fig.\ \ref{fig:expsetup}(a).

\begin{figure}[h]
	\includegraphics[width=\columnwidth]{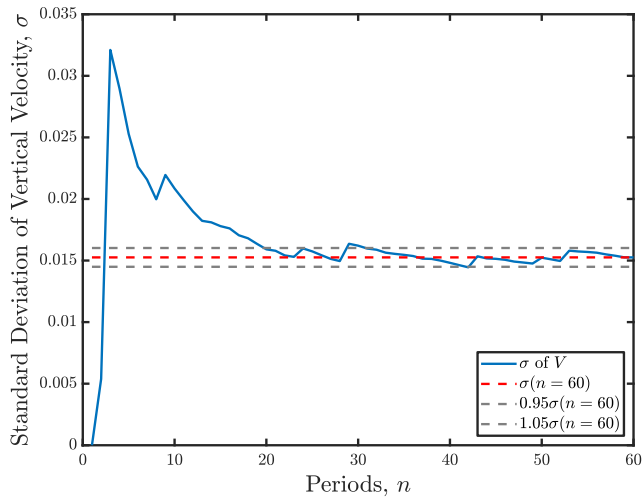}
    \caption{Standard deviation of the vertical velocity at $\frac{t}{T_0} = 0.8$ for $St = 0.096$ and $k = 0.603$, as a function of the number of test periods. Acceptable convergence is achieved by 30 periods.}
    \label{fig:stddev}
\end{figure}

In order to determine the number of test periods required to achieve statistical significance, the vertical velocity $V$ at one instance of time within a test period was averaged over 60 periods at the point described above. A protocol with a large plunge amplitude ($h_0/c = 0.5$) and zero pitch was used to ensure that dynamic effects would be present and would be incorporated into the analysis. The averaged values for each period were randomly shuffled, and a running value for the standard deviation $\sigma(n)$ was computed as a function of the number of test periods included in the computation. The results from this analysis, shown in Fig.\ \ref{fig:stddev}, determined that $n = 30$ test periods yielded reasonable convergence in the standard deviation (within $\pm 5\%$ of the final value, $\sigma(n=60)$). Thus, all data presented in this work are phase-averaged over at least 30 test periods. Furthermore, at the start of every experiment, the airfoil was oscillated through a few unrecorded motion periods to ensure that startup effects did not influence the phase-averaged data.

Finally, in order to quantify the two main parameters of interest for the gust-generation problem, two methods for analysis were developed. In order to quantify the sinusoidal character of the velocity fluctuations, the average acceleration at the theoretical peaks of the signal was measured using linear fits about $\frac{t}{T_0} \approx 0.25$ and $\frac{t}{T_0} \approx 0.75$. The dimensionless time $\frac{t}{T_0}$ was given by the total period $T_0$ and was phase-referenced to the sine function. The fits encompassed 10\% of $T_0$ on either side of the instant of interest. After correcting for sign, the slopes of these linear fits were averaged. For an ideal profile, this metric would yield a mean acceleration of $\frac{dV}{dt} = 0$ $\mathrm{m/s^2}$, indicating symmetry in the velocity profiles. The idea is illustrated in Fig.\ \ref{fig:accel} in the following section.

In order to quantify the influence of shed vortices in the wake of the airfoil on the gust region, the vertical extent of the wake vorticity was measured at the location of interest, $x = 1.0c$. For every time instant in the phase-averaged period, vectors of vorticity at $x = 1.0c$ and $y$ spanning the height of the domain were taken. These vectors were compiled into a 2D field in $y$ and $t$ to show the evolution of the thickness of the wake region over one period. The noisy vorticity fields were smoothed using a $3 \times 3$ filter, and were then binarized by an automatically adaptive threshold based on the standard deviation across the entire $y$-$t$ field. This isolated areas of relatively high vorticity. These fields were filtered using a 2D $7 \times 7$ median filter and remaining holes were filled to connect components within the identified region. The result of this procedure was a mask for the wake of the airfoil. The width of this mask was measured at every time instant across the phase-averaged cycle, and these values were then averaged to represent the mean width of the wake. Examples of vorticity profiles after application of the mask are shown in Fig.\ \ref{fig:wakewidth}. In the ideal case, this metric would be minimized, indicating minimally low interference of the gust generator itself on the flow downstream of the wake.

\section{Experimental Results}

In this section, the arguments put forward in Sec.\ \ref{sec:sec2_theodorsen} are examined experimentally. First, various experiments are shown to establish a baseline for effects of airfoil kinematics on the character of the gusts produced. Based on these baseline cases, a template for the kinematics of the gust generator is proposed, which agrees with the conclusions of the theoretical analysis. Lastly, the predictions of the Theodorsen theory as applied to gust generation are tested using the metrics outlined in Sec.\ \ref{sec:sec3_processing} in order to demonstrate that gusts with a satisfactorily sinusoidal character can be produced using a single pitching and plunging airfoil.

\subsection{Baseline cases}
\label{sec:sec4_baseline}

\begin{figure*}[!ht]
	\begin{subfigure}[t]{0.48\textwidth}
\centering
  \includegraphics[width=\textwidth]{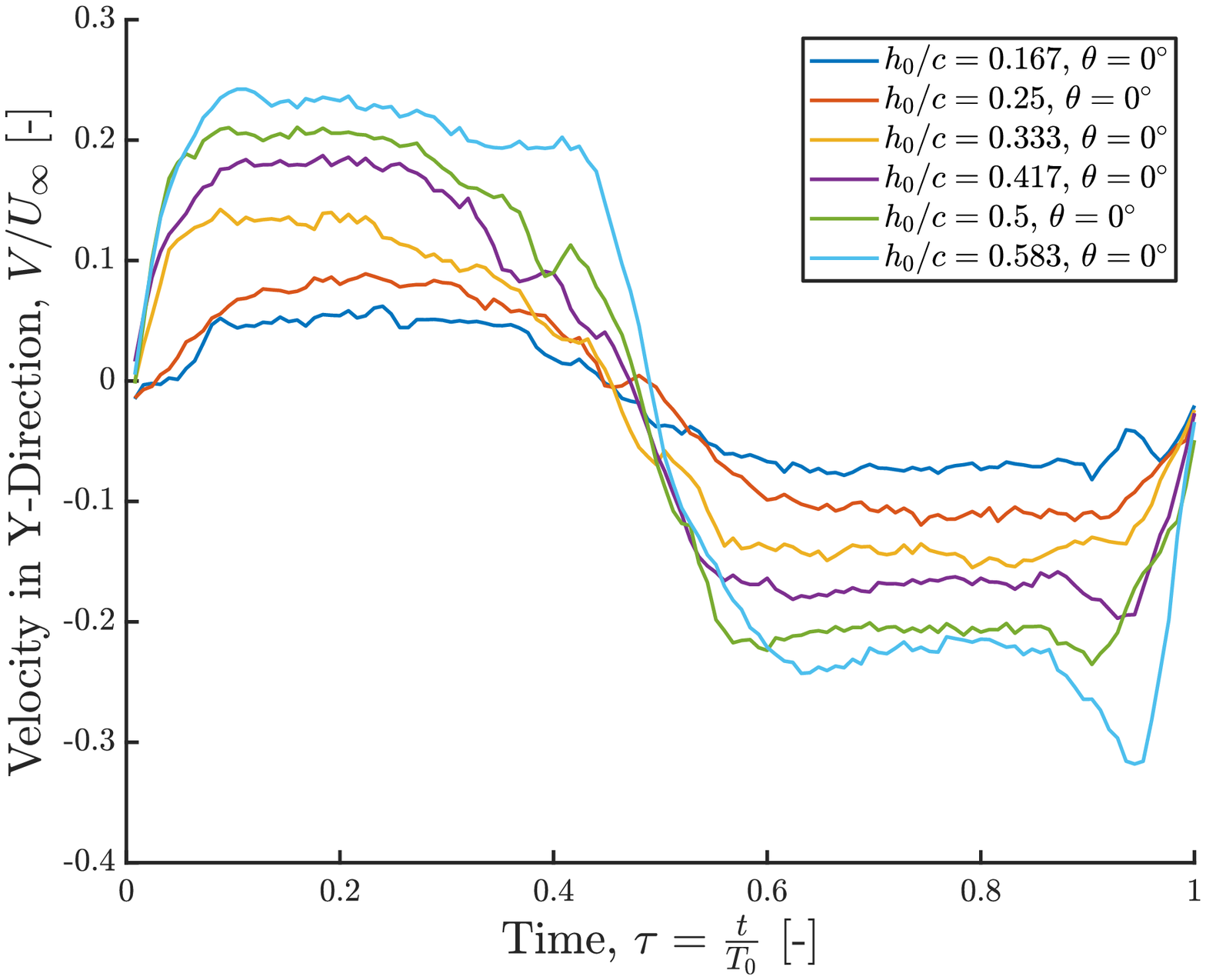}
  \caption{$0.032 \leq St \leq 0.112$ and $k = 0.603$}
\label{fig:baseline_St}
\end{subfigure}
\hfill
\begin{subfigure}[t]{0.48\textwidth}
\centering
  \includegraphics[width=\textwidth]{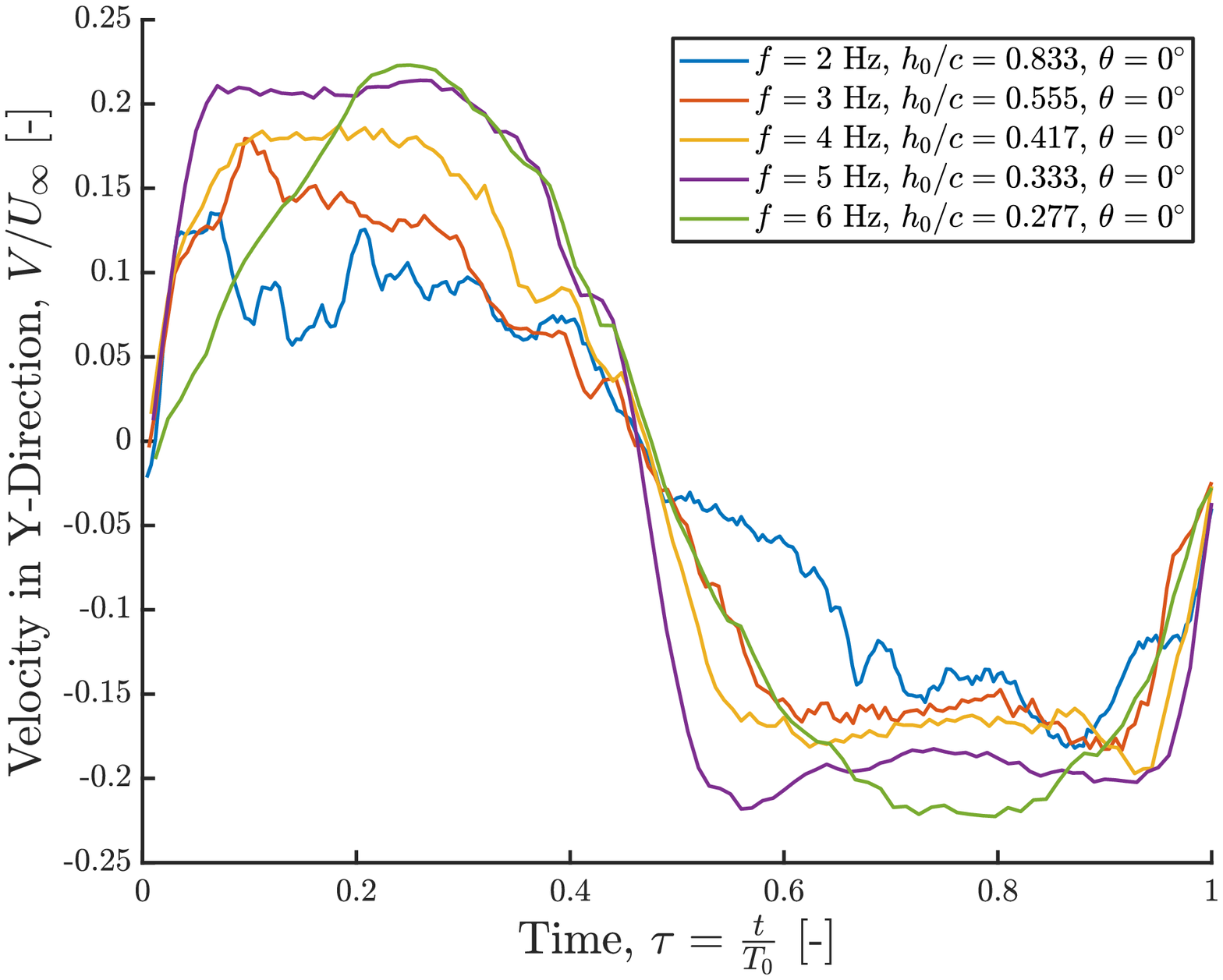}
  \caption{$0.302 \leq k \leq 0.905$ and $St = 0.080$}
\label{fig:baseline_k}
\end{subfigure}
    \caption{Baseline profiles of vertical velocity $V$ for a purely plunging airfoil for a range of (a) Strouhal numbers and (b) reduced frequencies. Measurements were taken one chord behind the airfoil's trailing edge, and were phase-averaged over 30 cycles. The peaks of the profiles tend to be biased toward earlier dimensionless times.}
    \label{fig:baselineStk}
\end{figure*}

To understand the effects of the various free parameters in the kinematics of a pitching and plunging airfoil on the vertical velocity signal in the wake, a series of concise parametric studies was conducted. First, the two dimensionless parameters governing the dynamics of the oscillating airfoil, $St$ and $k$, were varied independently for an airfoil moving in pure plunge. For $k = 0.603$, the plunge amplitude of the airfoil was varied to produce Strouhal numbers between $0.032$ and $0.112$. The resulting velocity profiles, taken at the region of interest specified in Sec.\ \ref{sec:sec3_processing}, are shown in Fig.\ \ref{fig:baseline_St}. Conversely, for $St = 0.080$, the reduced frequency was varied between $0.302$ and $0.905$. The velocity profiles for these experiments are shown in Fig.\ \ref{fig:baseline_k}. It is clear from these experiments that the amplitude of the fluctuations in $V$ was a strong linear function of Strouhal number and a weaker function of reduced frequency. In addition, most of the curves shown were asymmetric, with peaks occurring relatively early within each half-cycle, or more geometrically conceived, leftward-leaning maxima. This ``leaning" behavior should be avoided if quasi-sinusoidal signals are desired. Finally, in many cases, the waveform was interrupted by a deviation around $\frac{t}{T_0} = 0.95$ that was not present in the corresponding positive half of the signal. This was likely an artifact of flow around the actuator mounts interfering with the wake, an unavoidable asymmetry in the present experiment. Thus, though the negative portions of the waveforms show similar trends to the positive portions, the positive halves of the velocity profiles should be viewed as more reliable indicators of trends than the negative halves.

\begin{figure*}[!ht]
	\begin{subfigure}[t]{0.48\textwidth}
\centering
  \includegraphics[width=\textwidth]{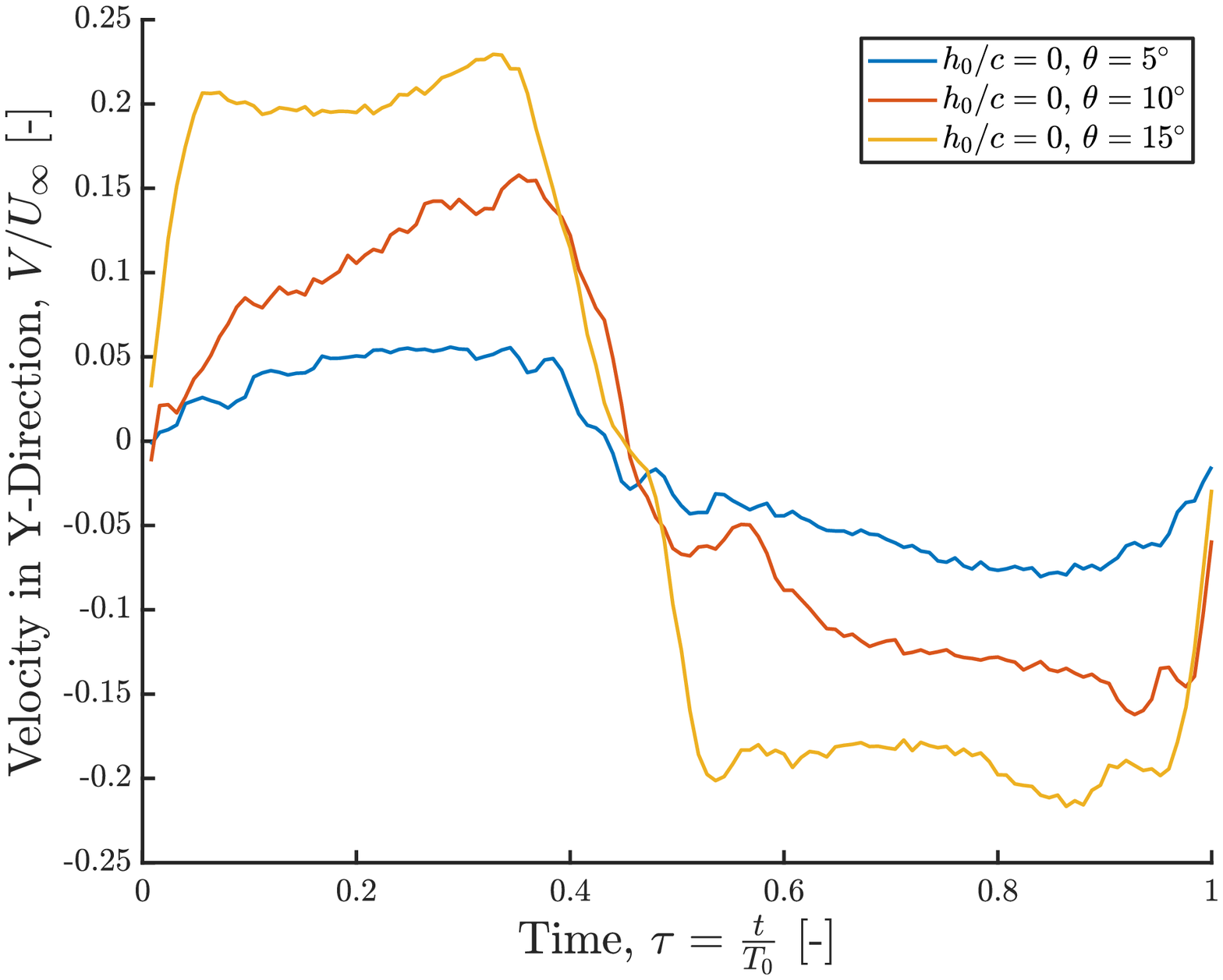}
  \caption{Pure pitch about the leading edge}
\label{fig:baseline_pitchLE}
\end{subfigure}
\hfill
\begin{subfigure}[t]{0.48\textwidth}
\centering
  \includegraphics[width=\textwidth]{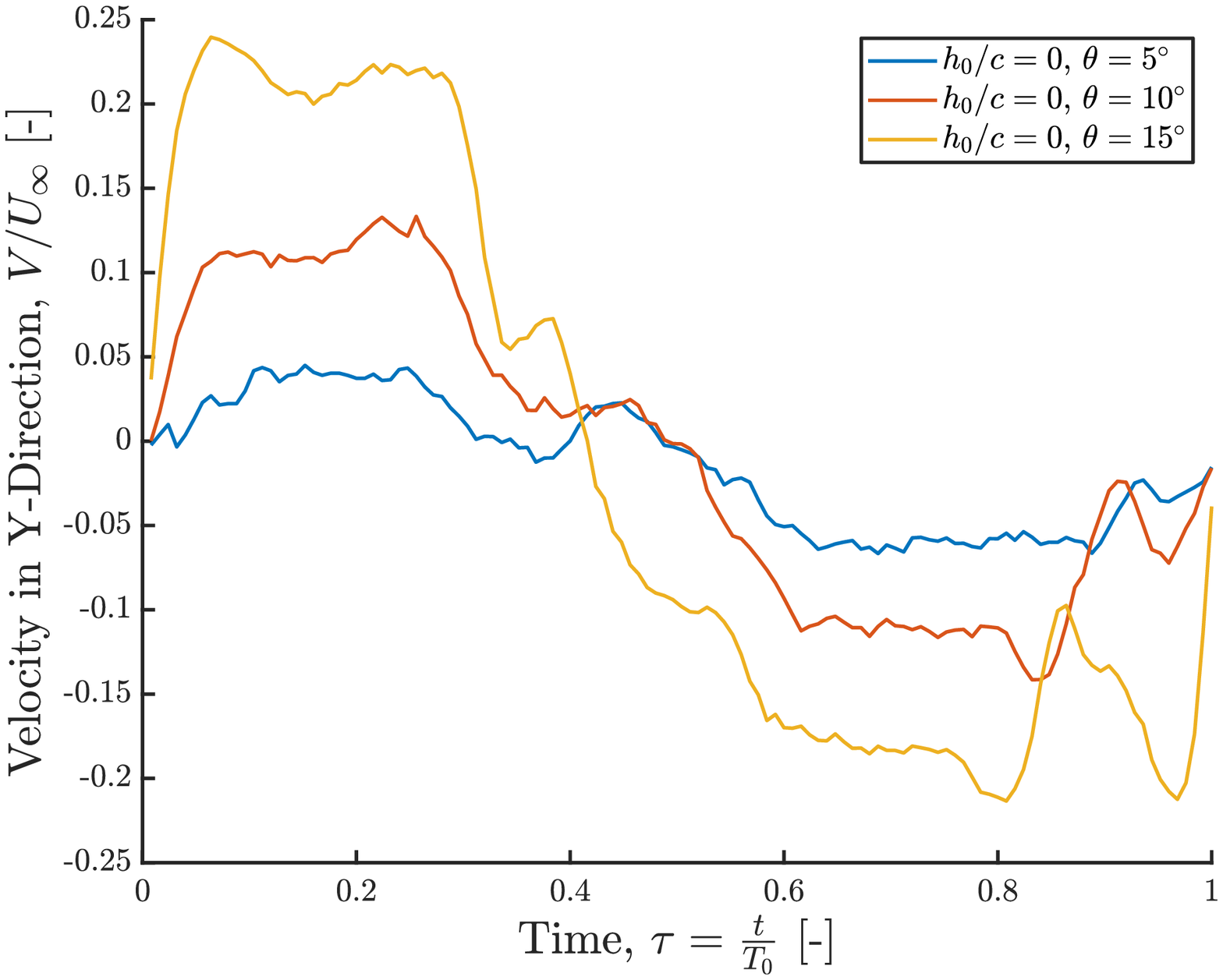}
  \caption{Pure pitch about the trailing edge}
\label{fig:baseline_pitchTE}
\end{subfigure}
    \caption{Baseline profiles of $V$ for a purely pitching airfoil ($k = 0.603$), actuated about its (a) leading edge and (b) trailing edge. The profiles appear to be more regular when leading-edge motion is minimized.}
    \label{fig:baselinePitchPoint}
\end{figure*}

Next, the effect of pure pitch on wake character was studied. The airfoil was pitched through $5^\circ \leq \theta \leq 15^\circ$, first about its leading edge and then about its trailing edge. The reduced frequency was again $k = 0.603$. The resulting profiles are shown in Figs.\ \ref{fig:baseline_pitchLE} and \ref{fig:baseline_pitchTE}. Upon inspection, pitching about the leading edge appeared to result in smoother, more regular profiles than pitching about the trailing edge. This is likely due to significant dynamic effects and resultant shedding of vortices that are instigated by large excursions of the leading edge of the airfoil \cite[cf.][]{widmann_parameters_2015}. Thus, it was concluded that excessive motions of the leading edge were counterproductive to the generation of smooth, well-defined velocity signals. Additionally, the velocity profiles in the case of pitch about the leading edge peaked later in each half-cycle. This suggested that the addition of pitch to plunge could potentially offset the leftward lean observed in Fig.\ \ref{fig:baselineStk}.

\begin{figure*}[!ht]
	\begin{subfigure}[t]{0.48\textwidth}
\centering
  \includegraphics[width=\textwidth]{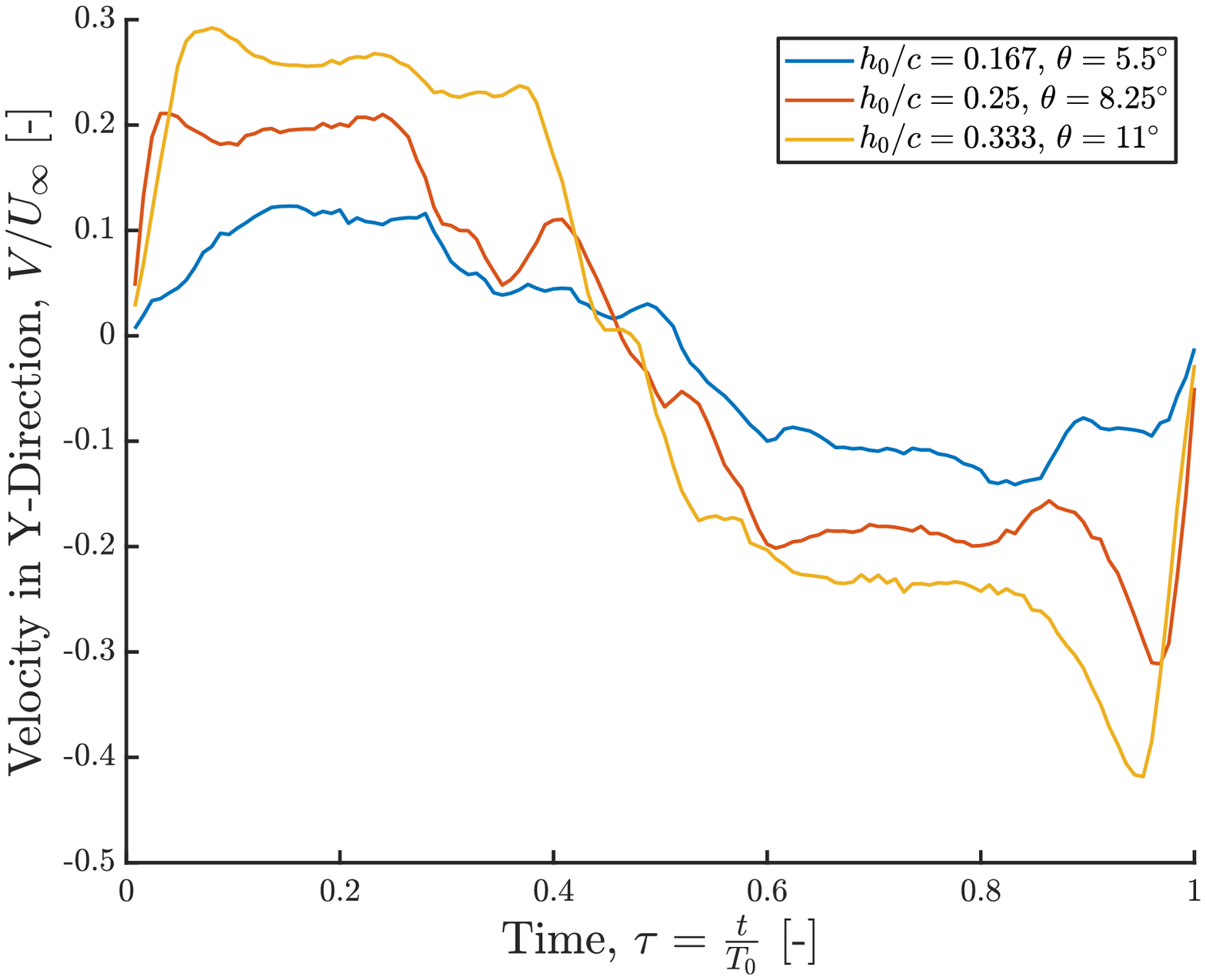}
  \caption{Phase 0}
\label{fig:baseline_phase0}
\end{subfigure}
\hfill
\begin{subfigure}[t]{0.48\textwidth}
\centering
  \includegraphics[width=\textwidth]{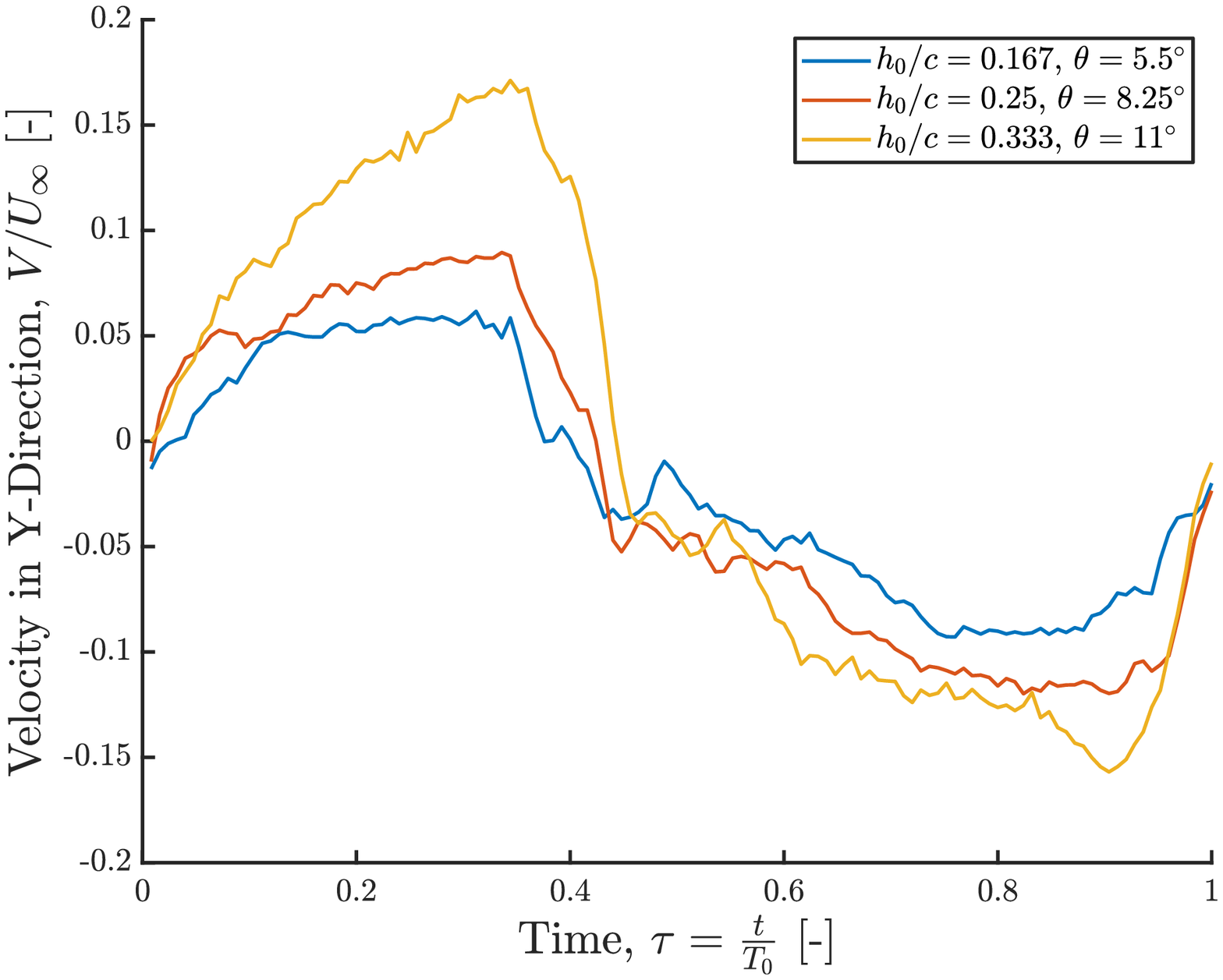}
  \caption{Phase 180}
\label{fig:baseline_phase180}
\end{subfigure}
    \caption{Baseline profiles of $V$ for a pitching and plunging airfoil ($k = 0.603$, range of $St$), with pitch phase (a) $\phi = 0^\circ$ and (b) $\phi = 180^\circ$. A phase of $180^\circ$ results in more regular profiles (due to the reduction in leading-edge motion) and biases the waveform peaks toward later dimensionless times.}
    \label{fig:baselinePitchPhase}
\end{figure*}

A similar conclusion was drawn by combining pitch and plunge kinematics, and changing the phase offset between the two. At $k = 0.603$, a set of three increasing pitch amplitudes was combined with a proportionally increasing set of plunge amplitudes. This series was carried out with phases of $\phi = 0^\circ$ and $180^\circ$. Since the trailing-edge amplitude defined the plunge waveform, a phase of $0^\circ$ produced larger excursions of the leading edge, and correspondingly less regular velocity profiles, as seen in Fig.\ \ref{fig:baseline_phase0}. Conversely, a phase of $180^\circ$ resulted in smaller motions of the leading edge; the velocity profiles (shown in Fig.\ \ref{fig:baseline_phase180}) exhibited the same rightward-shifted maxima as in Fig.\ \ref{fig:baseline_pitchLE}.

\begin{figure}[!ht]
	\includegraphics[width=\columnwidth]{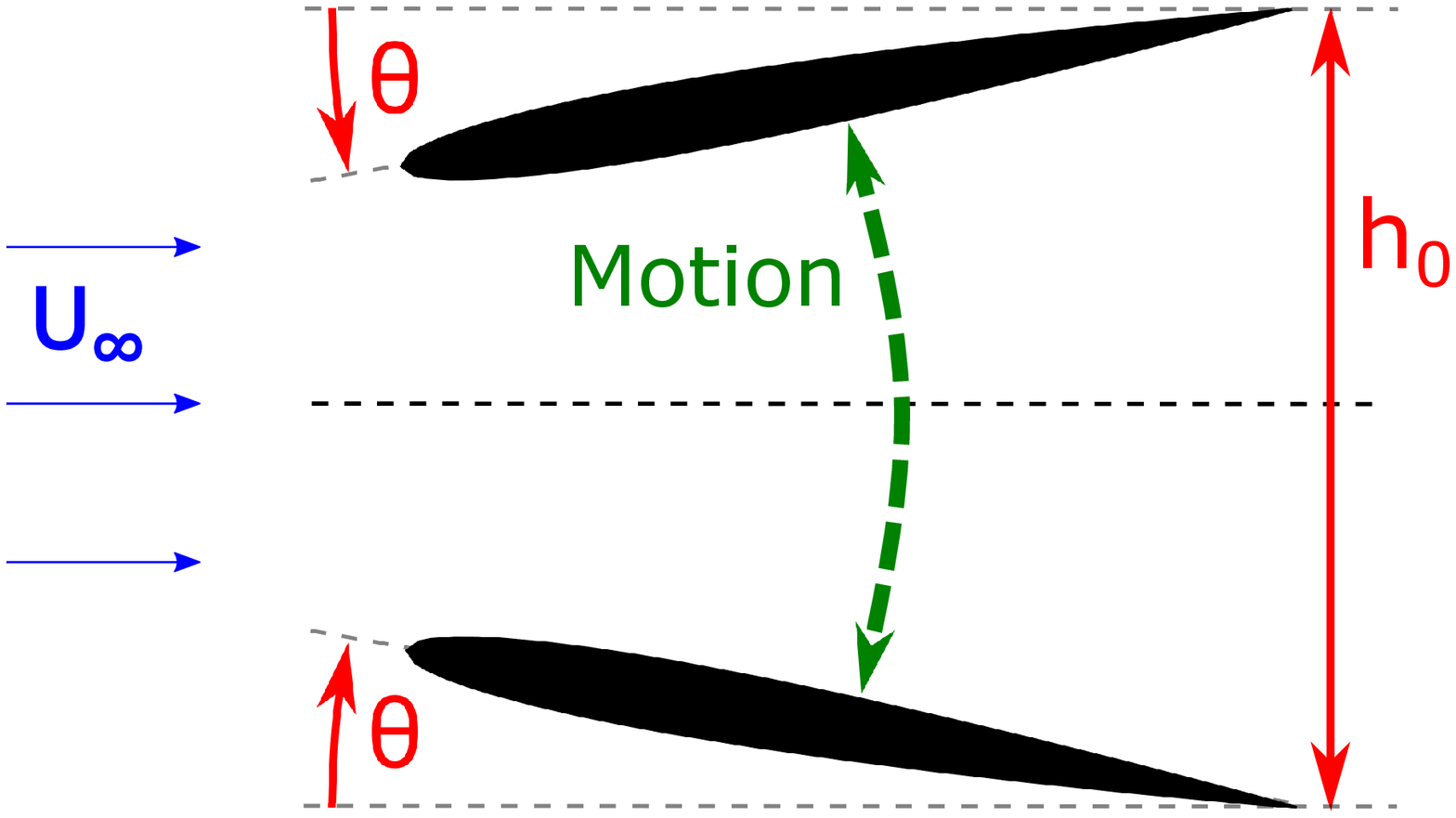}
    \caption{Schematic of the kinematics employed for gust generation, as determined by the Theodorsen theory and the set of baseline cases detailed above.}
    \label{fig:kinematics}
\end{figure}

The results observed in these small parametric studies suggested an ideal set of kinematics for the generation of sinusoidal gusts. Plunge motions allowed for higher gust amplitudes to be created, but introduced asymmetry into the gust profiles. Pitch created asymmetries in the opposite sense when significant leading-edge motion of the airfoil was avoided. A phase offset of $180^\circ$ between pitch and plunge, so that the leading edge of the airfoil moved through a smaller amplitude than the trailing edge, combined these considerations. This was thus selected as the ideal set of kinematics for sinusoidal-gust generation. A schematic of these kinematics is given as Fig.\ \ref{fig:kinematics}. These kinematics are identical to those identified in Sec.\ \ref{sec:sec2_theodorsen} by means of physical and analytical arguments.

\subsection{Generation of gusts with controlled character}
\label{sec:sec4_gustgen}

\begin{figure*}[!ht]
	\begin{subfigure}[t]{0.48\textwidth}
\centering
  \includegraphics[width=\textwidth]{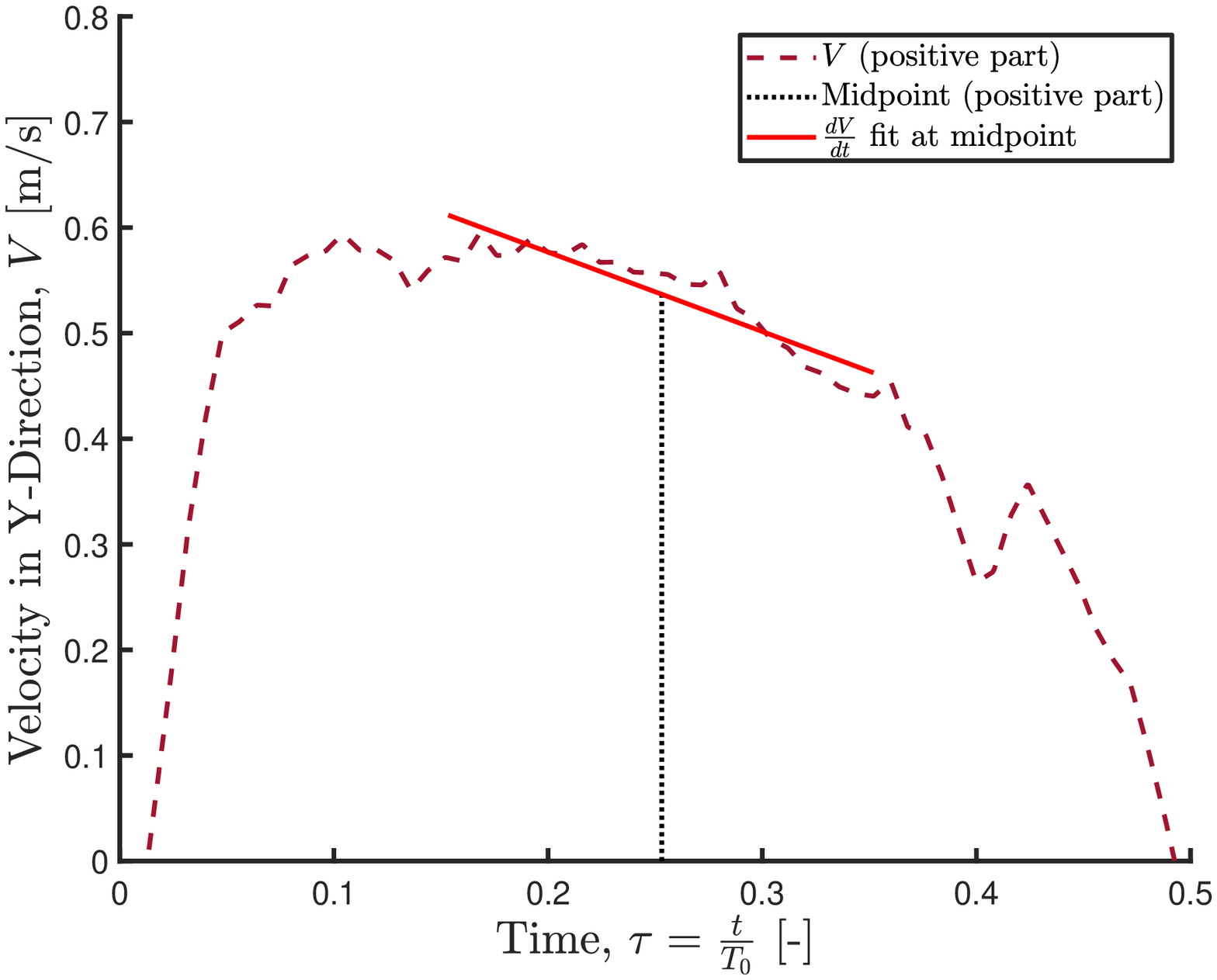}
  \caption{$\theta = 0^\circ$}
\label{fig:accelFits1}
\end{subfigure}
\hfill
\begin{subfigure}[t]{0.48\textwidth}
\centering
  \includegraphics[width=\textwidth]{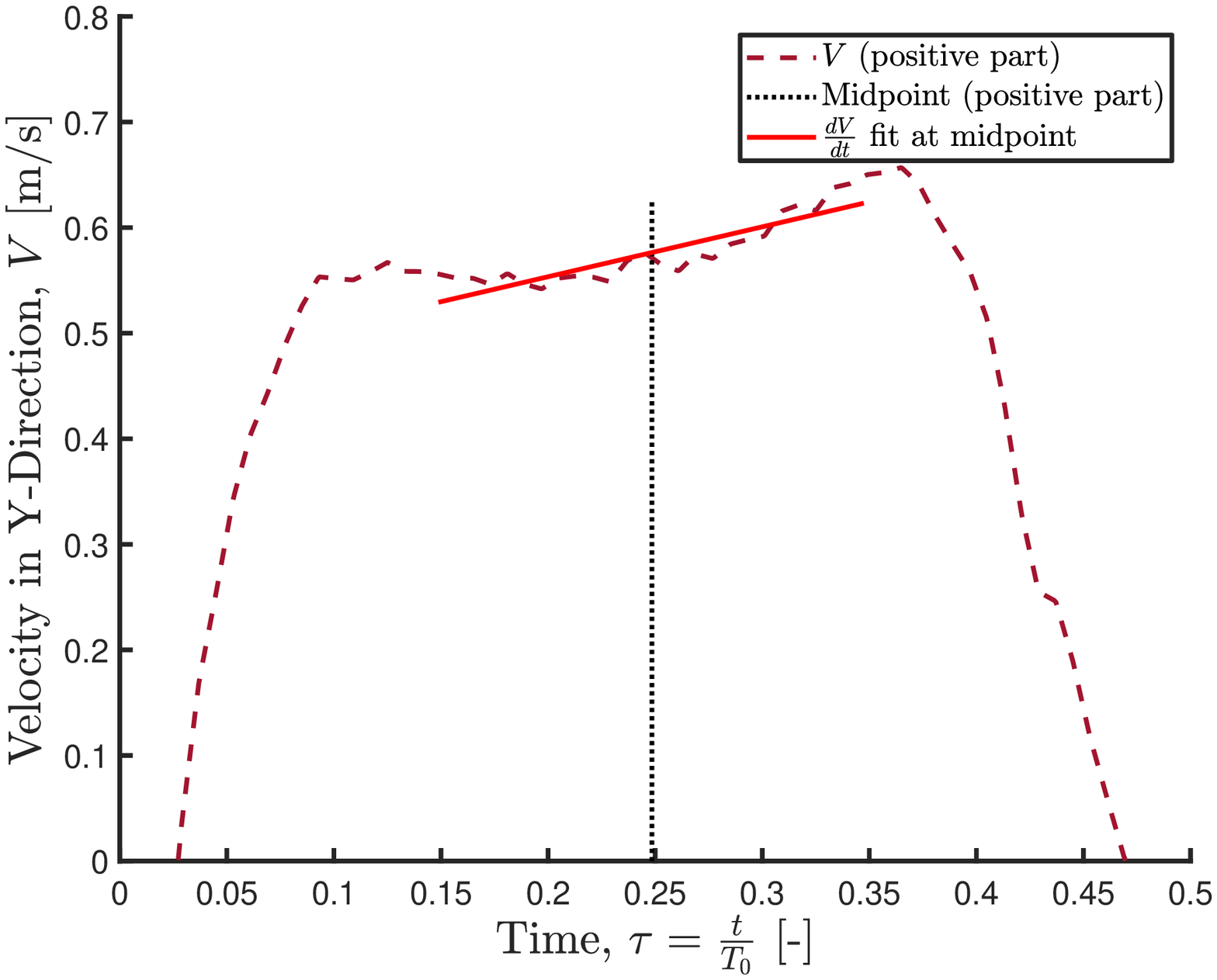}
  \caption{$\theta = 9.98^\circ$}
\label{fig:accelFits2}
\end{subfigure}
    \caption{Mean acceleration $\frac{dV}{dt}$ of the flow for the positive half of two velocity profiles from experiments with identical parameters save for pitch amplitude. The fit used to compute the mean acceleration is shown as a solid red line. The slope of the line quantifies the relative bias of the profiles, and thereby the deviation from the ideal, symmetric case.}
    \label{fig:accel}
\end{figure*}

\begin{figure*}[!ht]
	\includegraphics[width=\textwidth]{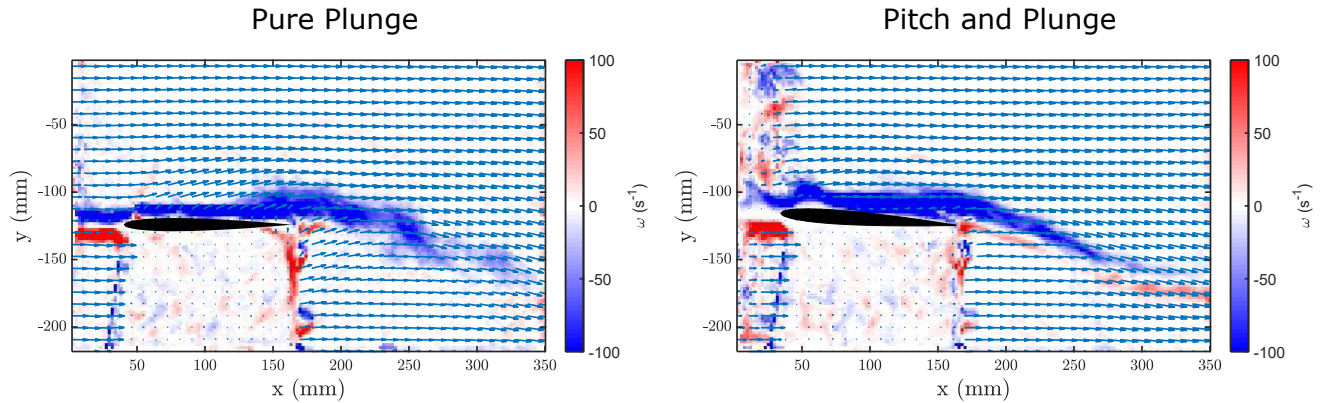}
    \caption{Vorticity fields (taken from Round 1 of experiments) for the same cases as shown in Fig.\ \ref{fig:accel}. Noise under the airfoil and on the left side of the frame in the rightmost figure is due to insufficient illumination by the laser sheet. The regularizing effect of pitch on the vorticity shed by the airfoil is evident in this comparison.}
    \label{fig:vorticity}
\end{figure*}

\begin{figure*}[!ht]
	\includegraphics[width=\textwidth]{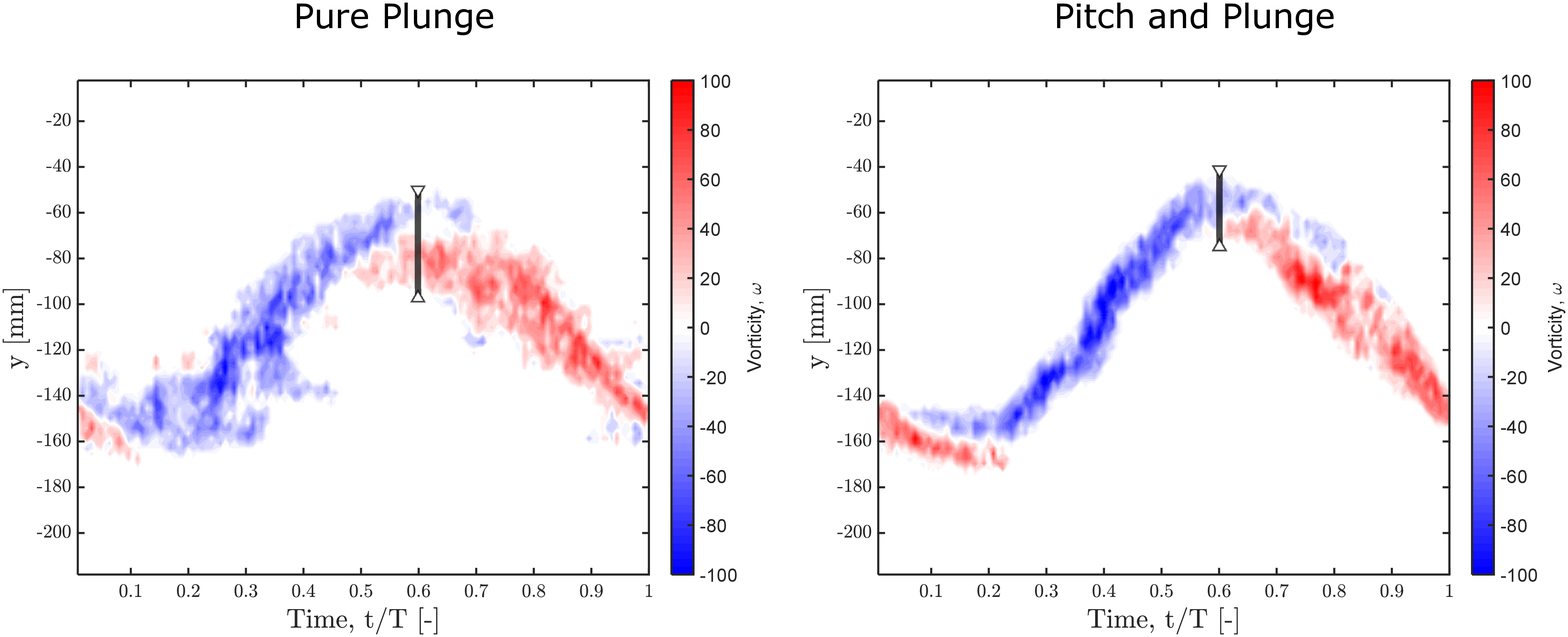}
    \caption{Phase-averaged wake profiles at $x = 1.0c$ behind the trailing edge of the airfoil, taken from the cases shown in Fig.\ \ref{fig:vorticity}. The wake widths are shown qualitatively by a gray vertical line at a single time instant in these profiles; the values of the wake width given in Fig.\ \ref{fig:optimizeWakeWidth} were averaged over the entire period.}
    \label{fig:wakewidth}
\end{figure*}

With the concurrence of theory and experiments in defining a set of kinematics for gust generation, it was then possible to test the specific predictions of the Theodorsen theory in relation to the gust-generation problem. First, a series of intermediate steps in the metrics described in Sec.\ \ref{sec:sec3_processing} are shown to demonstrate the effect of adding pitch out of phase with plunge. In Fig.\ \ref{fig:accel}, the positive half of the waveform for a case with $\theta = 0^\circ$ (Fig.\ \ref{fig:accelFits1}) is compared with the corresponding section for a case with $\theta = 9.98^\circ$. The addition of pitch reversed the sign of the average acceleration around the midpoint of the profile, further suggesting that, with the right pitch amplitude, the acceleration at the midpoint could be made to be zero. For the same cases, phase-averaged vorticity fields (shown in Fig.\ \ref{fig:vorticity}) demonstrated that the structure of the wake shed off the oscillating airfoil was thinner and more organized when pitch was present. This was also reflected in the wake-width analysis, representations of which are shown in Fig.\ \ref{fig:wakewidth}. These figures demonstrate the capacity of the metrics to capture physical differences in gust character within the area of interest, and also highlight the ability of phase-offset pitch to reverse some of the trends seen in the purely plunging cases.

Quantitative values from these two metrics were obtained for four combinations of Strouhal number and reduced frequency over a range of pitch amplitudes, in order to test the predictions of the Theodorsen theory. For a reduced frequency of $k = 0.603$, Strouhal numbers of $0.032$, $0.064$, and $0.096$ were tested. A case with $k = 0.754$ and $St = 0.080$ was added to ensure that the theory was valid across reduced frequencies as well. These cases will be denoted as (a), (b), (c), and (d), respectively. The airfoil kinematics for these cases and corresponding optimal pitch amplitudes according to theory ($\theta^*$) are given in Table \ref{tab:pitchAmpsOptimizedCases}.

\begin{table}[!ht]
\centering
\begin{tabular}{| p{0.8cm} | p{0.9cm} | p{1.2cm} | p{0.9cm} | p{0.9cm} | p{1cm} |}
\hline
Case & $f$ (Hz) & $h_0$ (mm) & $k$ & $St$ & $\theta^*$ \\ \hline\hline	
(a) & 4 & 20 & 0.603 & 0.032 & $2.06^\circ$ \\ \hline
(b) & 4 & 40 & 0.603 & 0.064 & $4.12^\circ$ \\ \hline
(c) & 4 & 60 & 0.603 & 0.096 & $6.18^\circ$ \\ \hline
(d) & 5 & 40 & 0.754 & 0.080 & $5.74^\circ$ \\ \hline
\end{tabular}
\caption{Kinematic parameters and corresponding optimal pitch amplitudes $\theta^*$ for the four cases shown in Figs.\ \ref{fig:optimizeAccel} and \ref{fig:optimizeWakeWidth}. For all cases, $U_\infty = 2.5$ m/s.}
\label{tab:pitchAmpsOptimizedCases}
\end{table}

\begin{figure*}[!ht]
	\begin{subfigure}[t]{0.40\textwidth}
\centering
  \includegraphics[width=\textwidth]{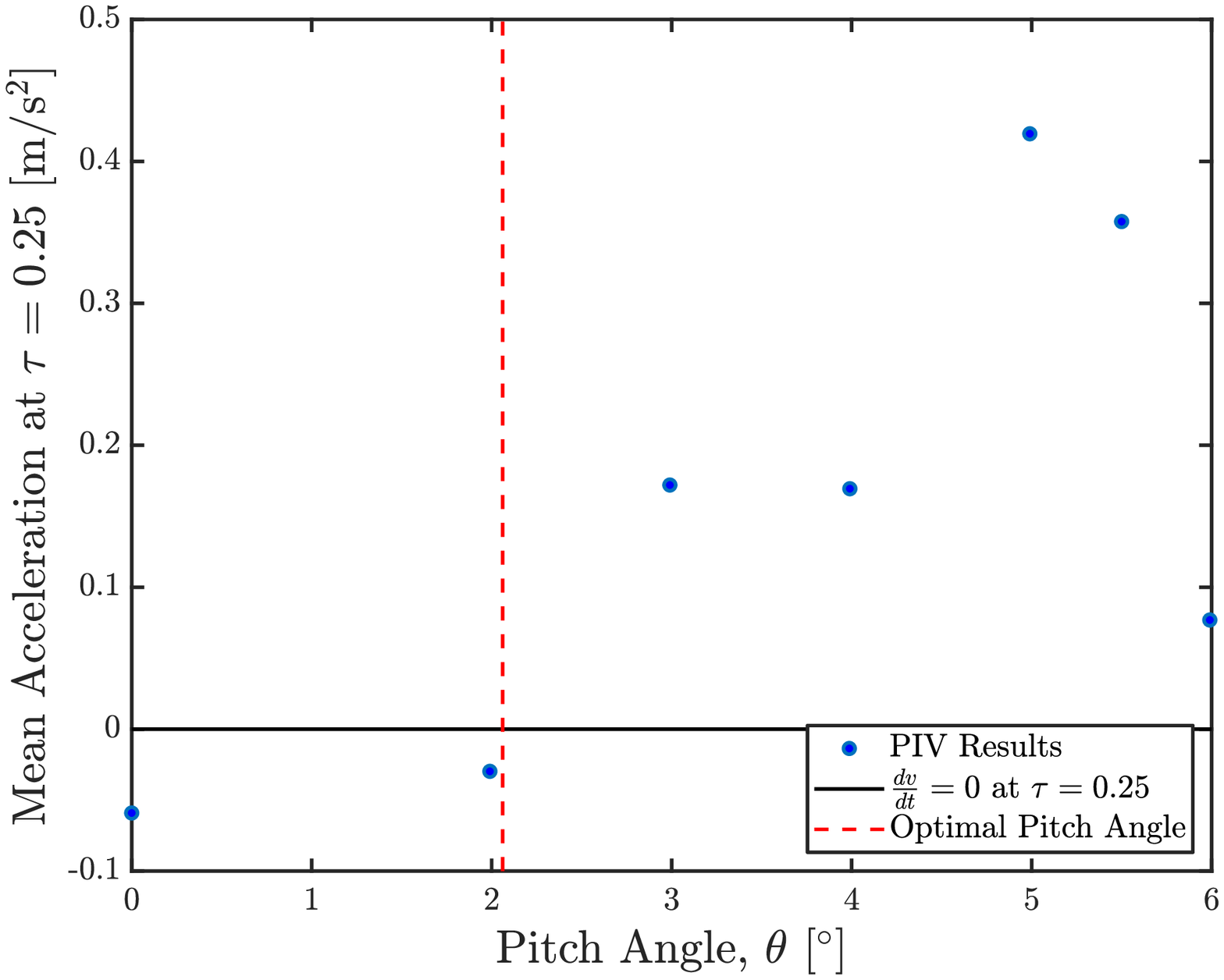}
  \caption{$St = 0.032$, $k = 0.603$}
\label{fig:accel1}
\end{subfigure}
\hfill
\begin{subfigure}[t]{0.40\textwidth}
\centering
  \includegraphics[width=\textwidth]{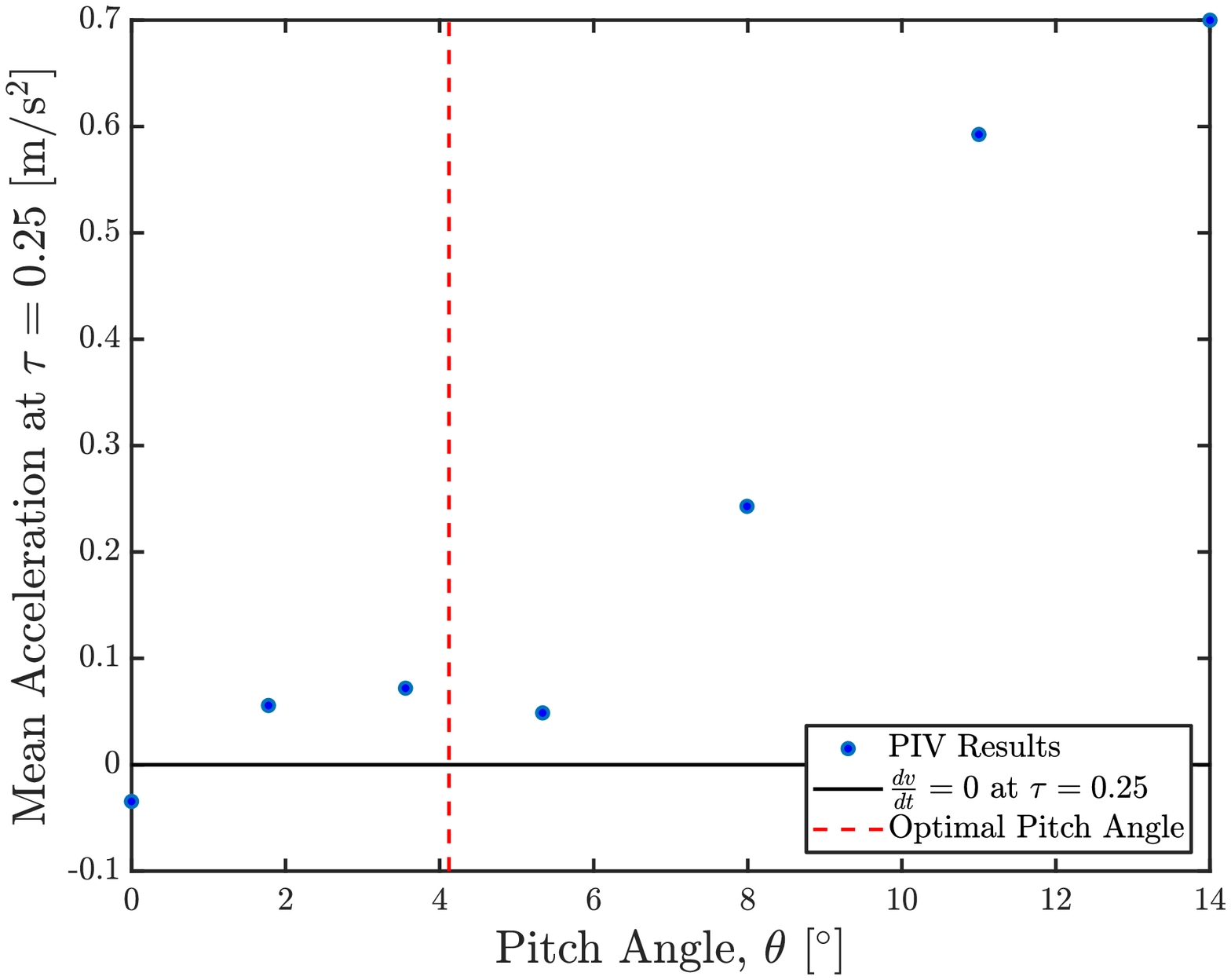}
  \caption{$St = 0.064$, $k = 0.603$}
\label{fig:accel2}
\end{subfigure}
\begin{subfigure}[t]{0.40\textwidth}
\centering
  \includegraphics[width=\textwidth]{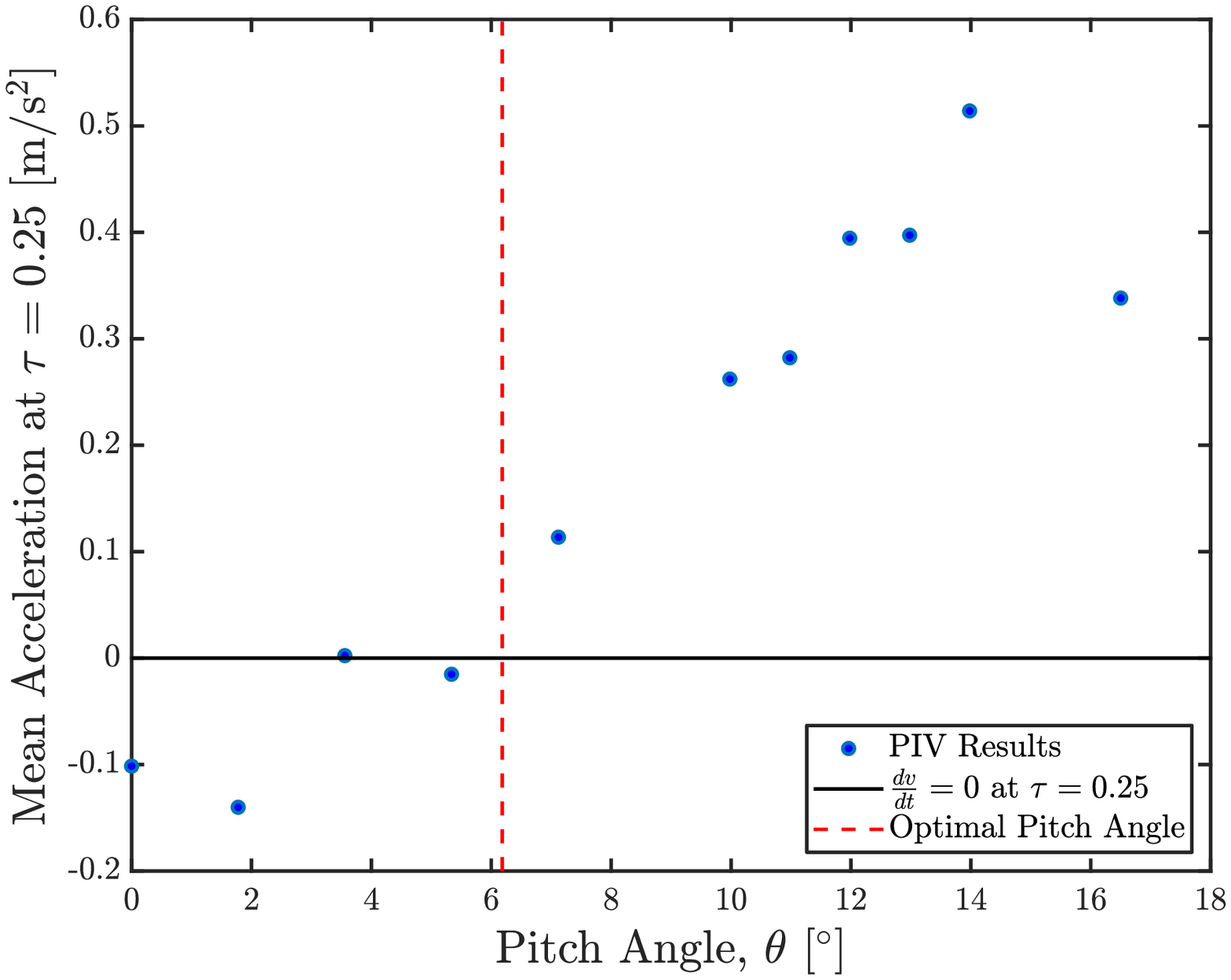}
  \caption{$St = 0.096$, $k = 0.603$}
\label{fig:accel3}
\end{subfigure}
\hfill
\begin{subfigure}[t]{0.40\textwidth}
\centering
  \includegraphics[width=\textwidth]{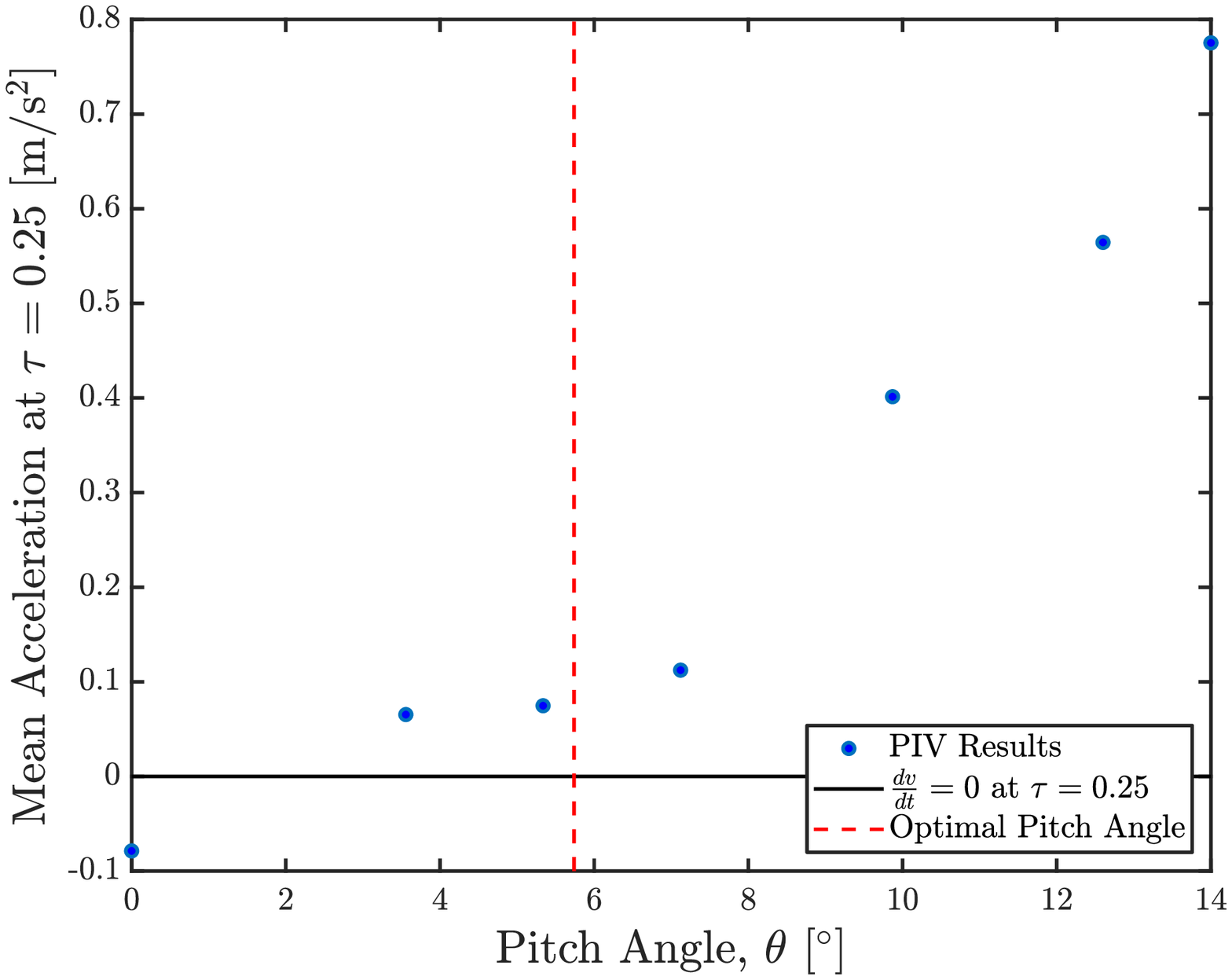}
  \caption{$St = 0.080$, $k = 0.754$}
\label{fig:accel4}
\end{subfigure}
    \caption{Mean acceleration $\frac{dV}{dt}$ of the flow at $\frac{t}{T_0} = 0.25$ and $\frac{t}{T_0} = 0.75$ for four combinations of Strouhal number and reduced frequency. As shown by the intersection of the data trends with $\frac{dV}{dt} = 0$ at the red vertical dashed lines (denoting $\theta^*$), the theory satisfactorily predicts the generation of optimally symmetric gust profiles.}
    \label{fig:optimizeAccel}
\end{figure*}

Fig.\ \ref{fig:optimizeAccel} shows the mean acceleration data for each of these cases over a range of pitch amplitudes. The prediction of the Theodorsen theory, shown as a red vertical dashed line on each plot, coincides very well with the point where the trend of the data crosses the $\frac{dV}{dt}$ axis in each case, in spite of minor scatter in the data due to experimental error stemming from the limits of the apparatus and the sensitivity of the mean-acceleration metric.

\begin{figure*}[!ht]
	\begin{subfigure}[t]{0.40\textwidth}
\centering
  \includegraphics[width=\textwidth]{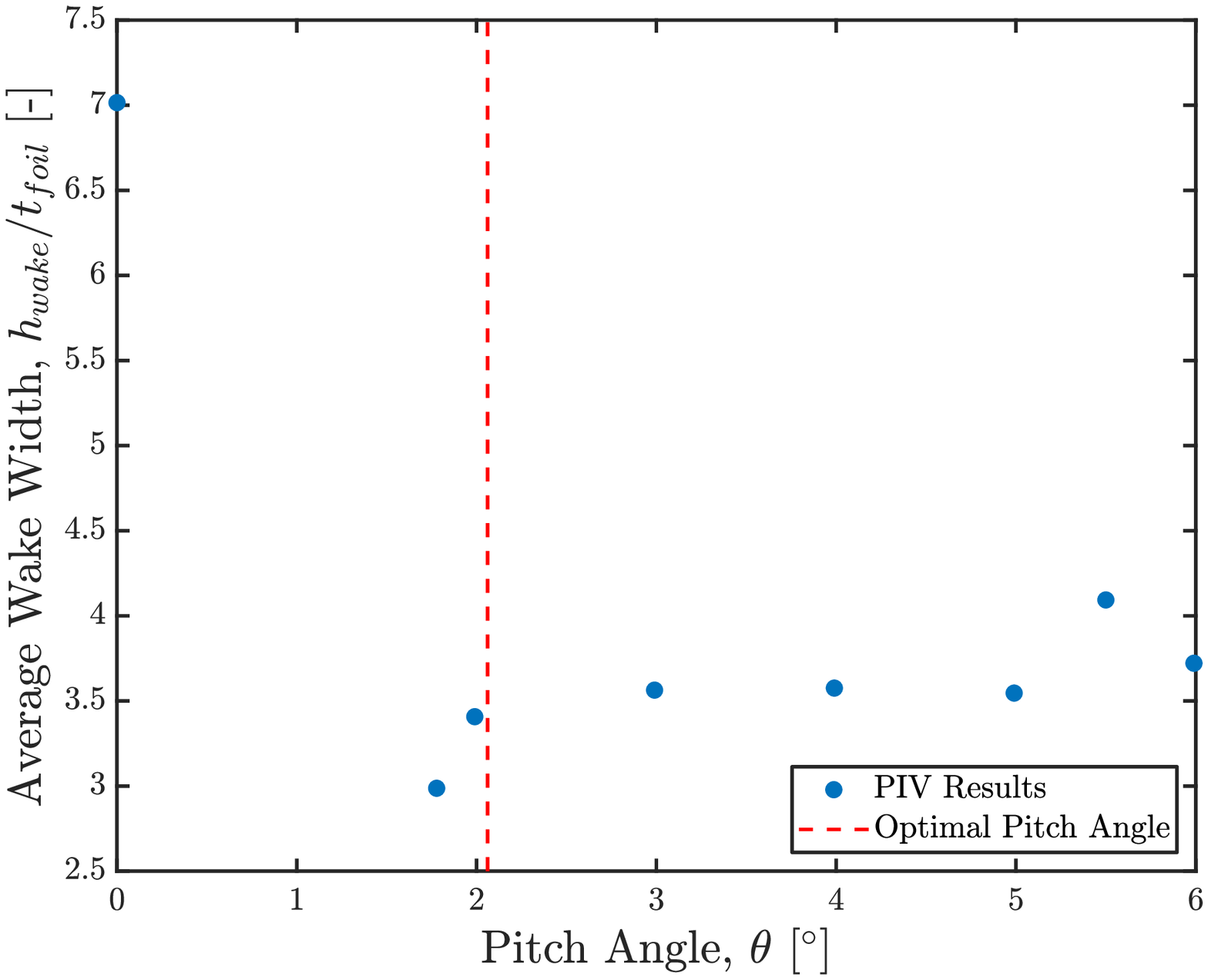}
  \caption{$St = 0.032$, $k = 0.603$}
\label{fig:wakewidth1}
\end{subfigure}
\hfill
\begin{subfigure}[t]{0.40\textwidth}
\centering
  \includegraphics[width=\textwidth]{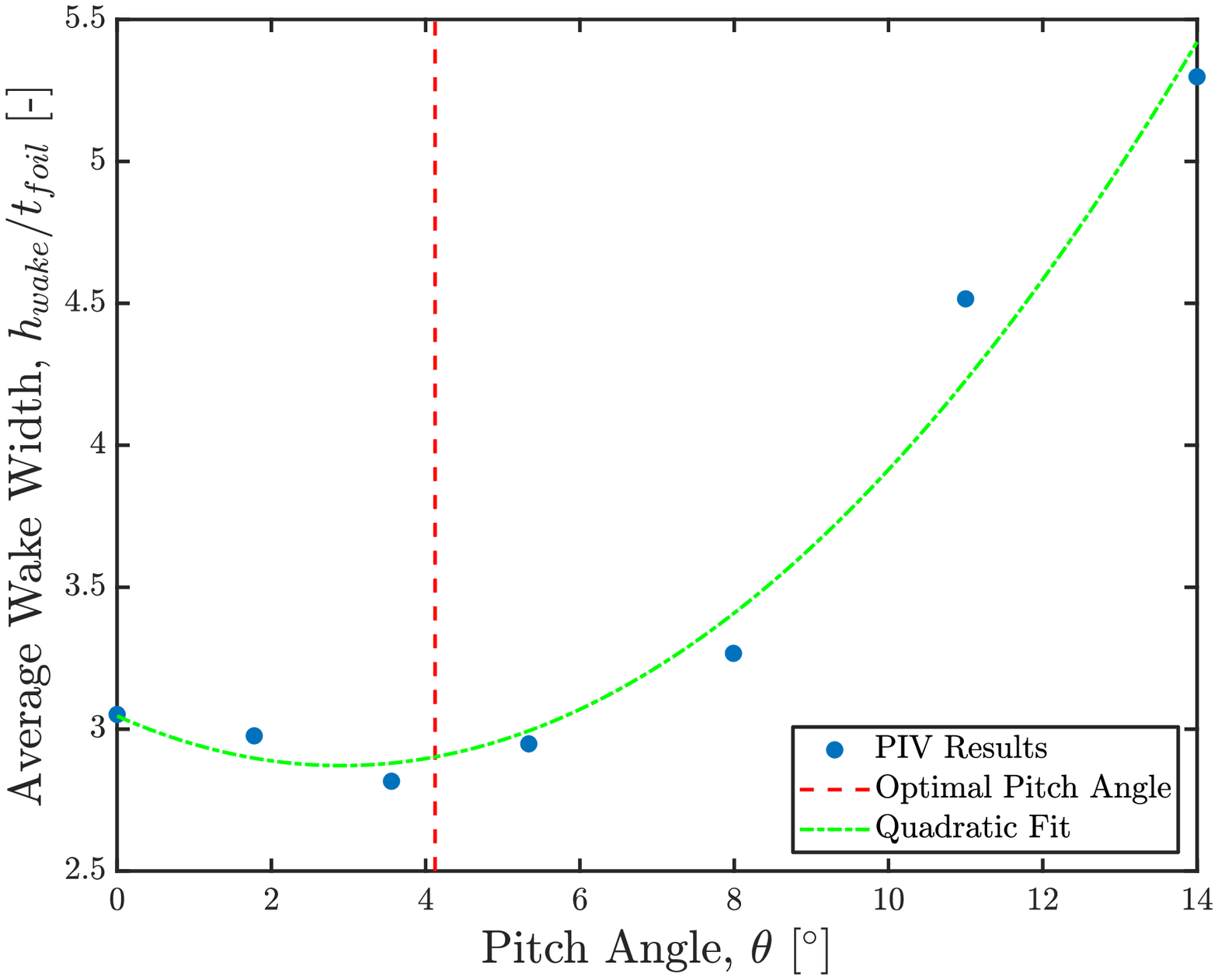}
  \caption{$St = 0.064$, $k = 0.603$}
\label{fig:wakewidth2}
\end{subfigure}
\begin{subfigure}[t]{0.40\textwidth}
\centering
  \includegraphics[width=\textwidth]{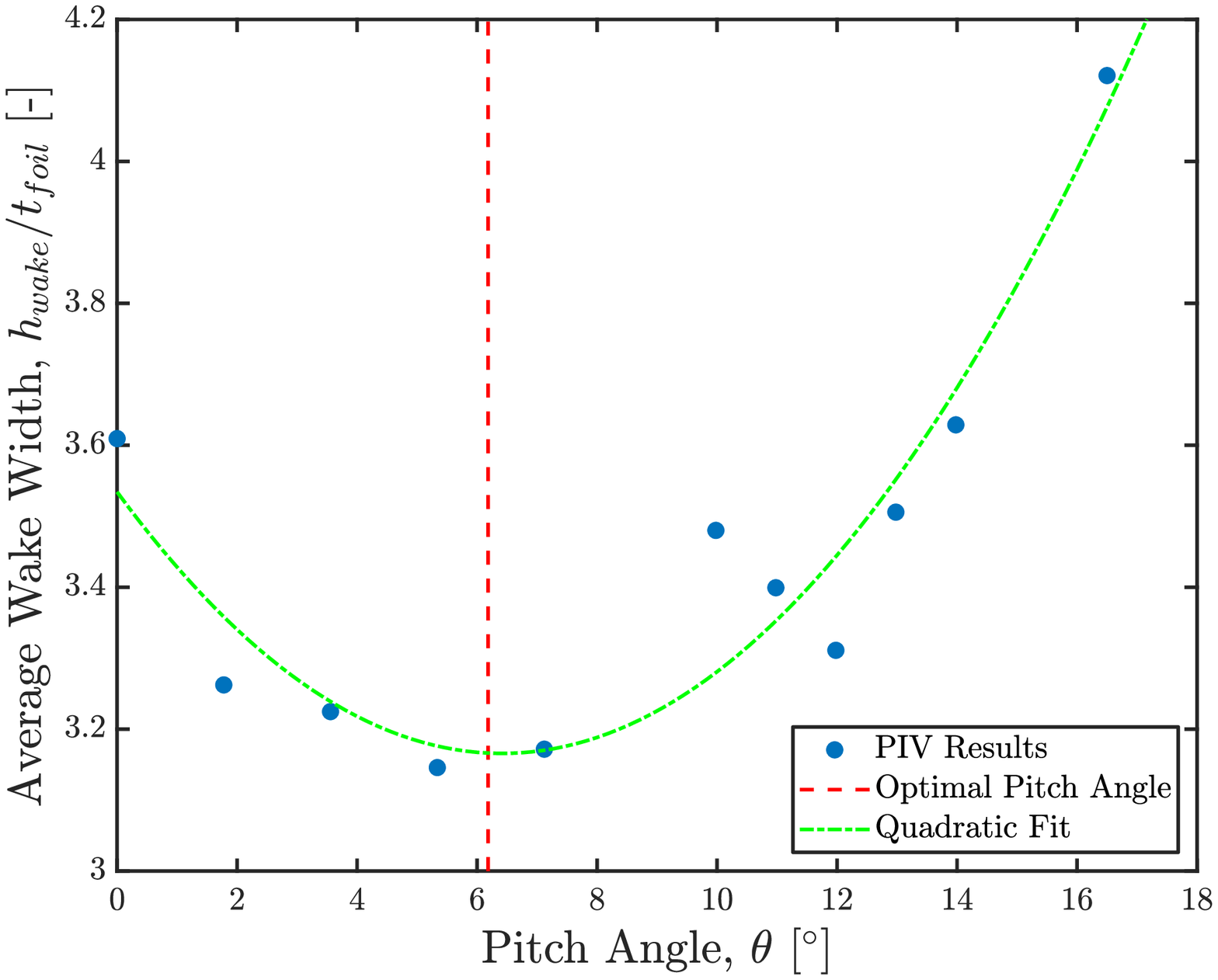}
  \caption{$St = 0.096$, $k = 0.603$}
\label{fig:wakewidth3}
\end{subfigure}
\hfill
\begin{subfigure}[t]{0.40\textwidth}
\centering
  \includegraphics[width=\textwidth]{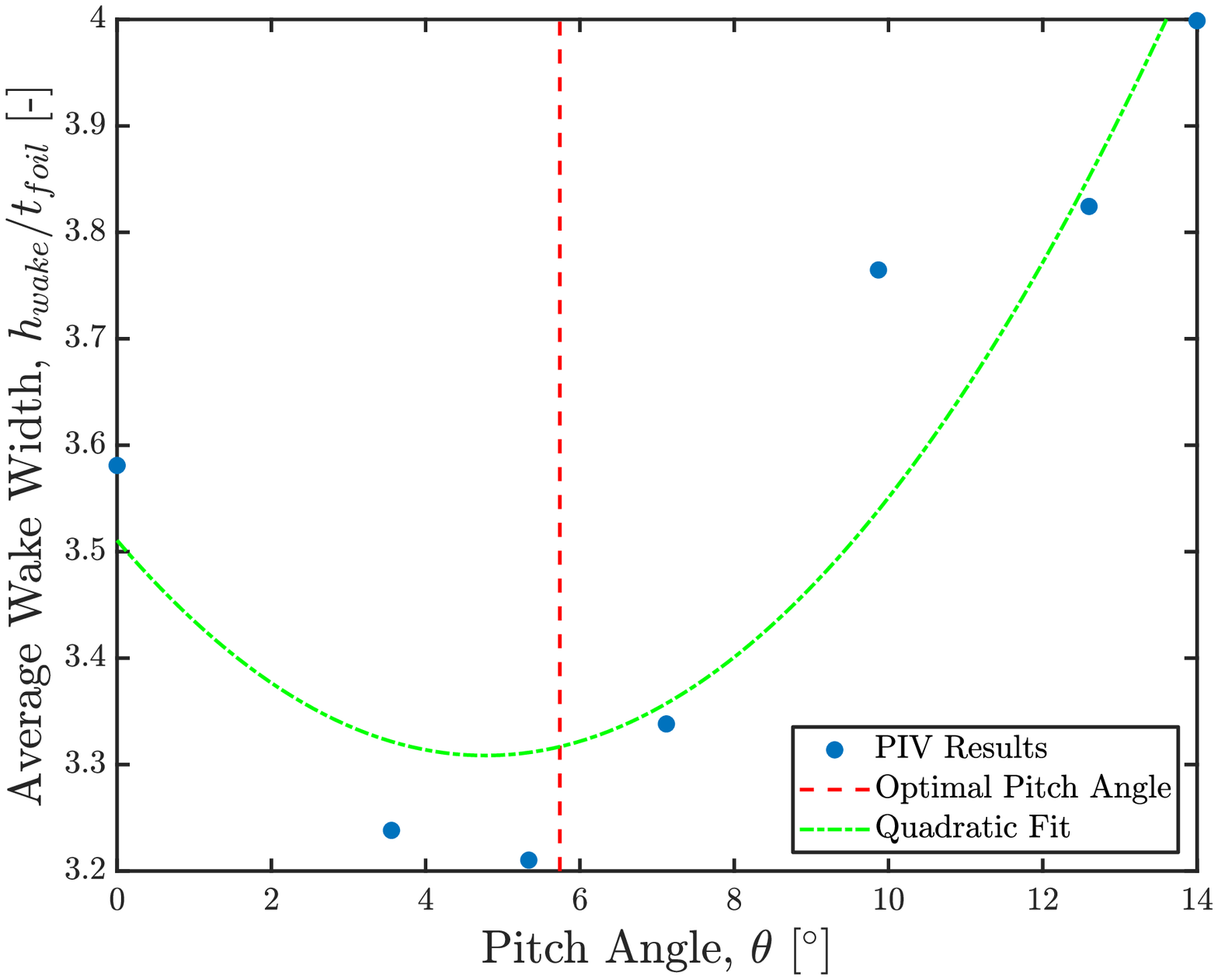}
  \caption{$St = 0.080$, $k = 0.754$}
\label{fig:wakewidth4}
\end{subfigure}
    \caption{Wake width, normalized by airfoil thickness $t_{foil}$, of the flow at $x = 1.0c$ for four combinations of Strouhal number and reduced frequency. Parabolic fits are given as dashed curves to show the trends, except for case (a). The influence of the wake of the gust generator is minimized in all cases near the pitch amplitude $\theta^*$ prescribed by the Theodorsen theory.}
    \label{fig:optimizeWakeWidth}
\end{figure*}

Similarly, the wake width (normalized by the thickness of the airfoil, $t_{foil}$) also agrees well with the predictions of the Theodorsen theory. In all four cases, shown in Fig.\ \ref{fig:optimizeWakeWidth}, a relative minimum in wake width was observed very close to the optimal pitch amplitude predicted by the Theodorsen analysis. The parabolic fits are only shown as guides for the eye, and are not intended to be prescriptive in any way. The effects of error in these data are more evident, stemming from the inherent noise associated with computing vorticity from velocity fields, the sensitivity of the thresholding procedure used to compute the wake width, and the fact that the field of view was so large (with respect to limits on laser power and seeding density) that achieving both sufficient illumination and sufficient resolution were difficult. Case (a), though still showing a minimum at the value of $\theta^*$ given by the Theodorsen theory, is less conclusive because of its low plunge amplitude. Inspection of the vorticity fields shows that the excursion of the airfoil, and thus the extent of the gust, is only marginally larger than the wake of the airfoil in the stationary case. As the flow in this regime was dominated by Kelvin-Helmholtz vortices from the surface of the airfoil rather than the unsteadiness of the airfoil, differences in the gust character were therefore much harder to measure. In spite of the influence of noise, every case investigated demonstrated general agreement with the Theodorsen theory in terms of a reduced influence of the airfoil in the downstream wake.

\begin{figure*}[!ht]
	\begin{subfigure}[t]{0.48\textwidth}
\centering
  \includegraphics[width=\textwidth]{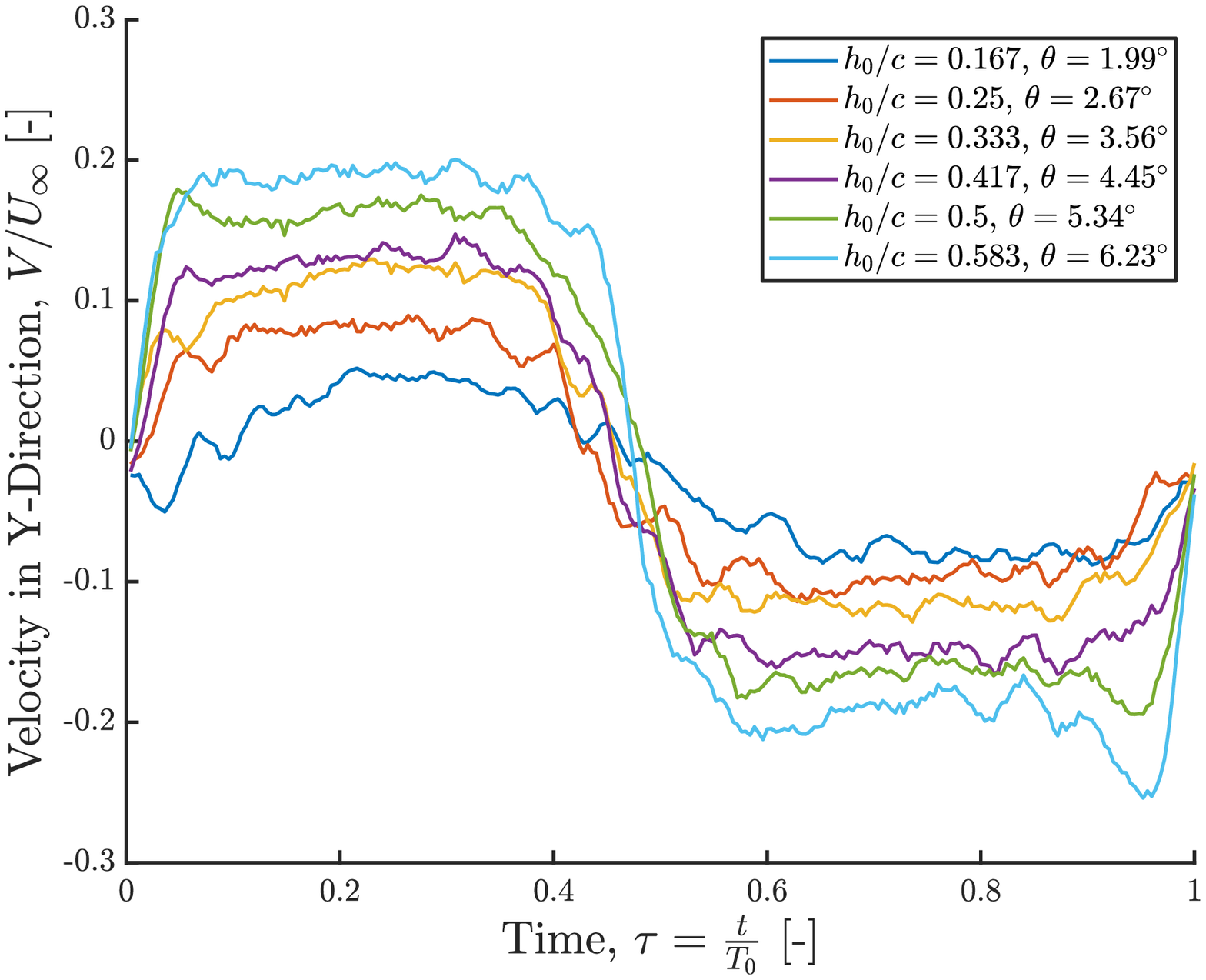}
  \caption{$0.032 \leq St \leq 0.112$ and $k = 0.603$}
\label{fig:optimized_St}
\end{subfigure}
\hfill
\begin{subfigure}[t]{0.48\textwidth}
\centering
  \includegraphics[width=\textwidth]{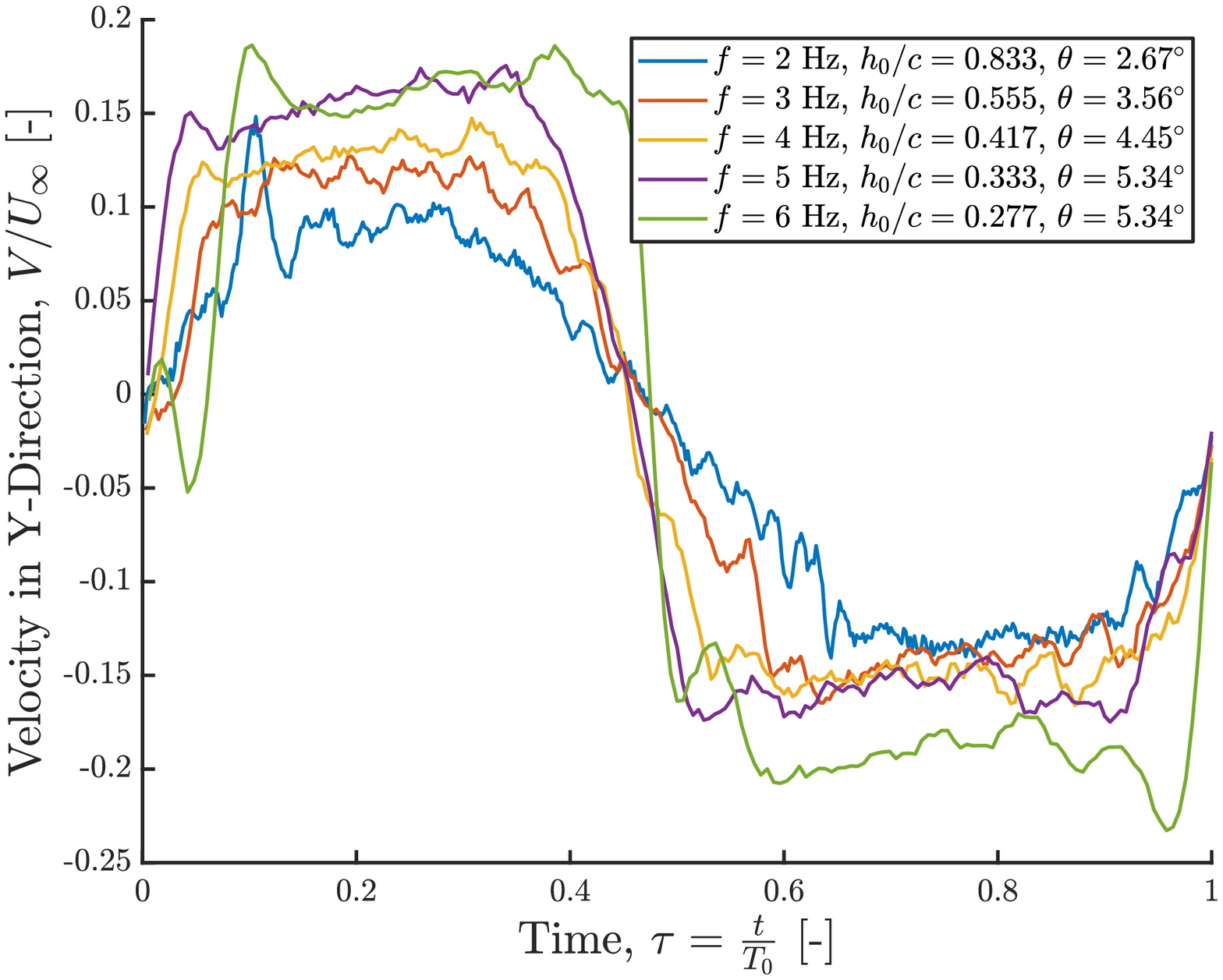}
  \caption{$0.302 \leq k \leq 0.905$ and $St = 0.080$}
\label{fig:optimized_k}
\end{subfigure}
    \caption{Profiles of $V$ behind a pitching and plunging airfoil for a range of (a) Strouhal numbers and (b) reduced frequencies. Despite the actual pitch amplitudes being slightly lower than those prescribed by the theory (due to a software error), these curves still compare favorably in terms of sinusoidal character to those in Fig.\ \ref{fig:baselineStk}.}
    \label{fig:optimizedStk}
\end{figure*}

Finally, in order to more clearly demonstrate the effectiveness of the Theodorsen theory for gust generation, a range of Strouhal numbers and reduced frequencies were tested with the addition of pitch according to the theory. The values of these optimal pitch amplitudes are given in Table \ref{tab:pitchAmpsBaselineCases}. These results could then be compared directly with the baseline cases shown in Fig.\ \ref{fig:baselineStk}. A small experimental oversight meant that the pitch amplitudes tested in the experiments were not precisely equivalent to those supplied by the theory; however, the gust profiles -- shown in Fig.\ \ref{fig:optimizedStk} -- were still much improved in comparison to the baseline cases. The leftward shift of the maxima visible in Fig.\ \ref{fig:baseline_St} was essentially removed for all cases in Fig.\ \ref{fig:optimized_St}. At higher reduced frequencies, shown in Fig.\ \ref{fig:optimized_k}, a slight rightward bias became evident. This was likely due to the onset of dynamic effects that, at high enough reduced frequencies, ceased to be mitigated by kinematics alone and began to significantly influence the wake. These deviations are unavoidable, as the Theodorsen theory is not able to account for significant dynamic-stall events. Overall, the addition of pitch according to theory regularized the vertical-velocity profiles over a range of Strouhal numbers and reduced frequencies, making the resultant gusts more uniform in shape and less affected by the wake of the airfoil generating the gust. These gusts were also significantly higher in amplitude (up to $11^\circ$) and reduced frequency (up to $k = 0.905$) than those produced by generators with the same number of actuators \cite[e.g.][]{lancelot_design_2015, wood_new_2017}, as well as more complex systems with multiple vanes \cite[e.g.][]{cordes_note_2017}.

\begin{table}[!ht]
\centering
\begin{tabular}{| p{1.1cm} | p{0.9cm} | p{1.2cm} | p{0.8cm} | p{0.8cm} | p{0.8cm} |}
\hline
Variable & $f$ (Hz) & $h_0$ (mm) & $k$ & $St$ & $\theta^*$ \\ \hline\hline	
$St$ & 4 & 20 & 0.603 & 0.032 & $2.06^\circ$ \\ \hline
$St$ & 4 & 30 & 0.603 & 0.048 & $3.09^\circ$ \\ \hline
$St$ & 4 & 40 & 0.603 & 0.064 & $4.12^\circ$ \\ \hline
$St$ & 4 & 50 & 0.603 & 0.080 & $5.15^\circ$ \\ \hline
$St$ & 4 & 60 & 0.603 & 0.096 & $6.18^\circ$ \\ \hline
$St$ & 4 & 70 & 0.603 & 0.112 & $7.21^\circ$ \\ \hline\hline
$k$ & 2 & 100 & 0.302 & 0.080 & $2.98^\circ$ \\ \hline
$k$ & 3 & 66.7 & 0.452 & 0.080 & $4.23^\circ$ \\ \hline
$k$ & 4 & 50 & 0.603 & 0.080 & $5.15^\circ$ \\ \hline
$k$ & 5 & 40 & 0.754 & 0.080 & $5.74^\circ$ \\ \hline
$k$ & 6 & 33.3 & 0.905 & 0.080 & $6.04^\circ$ \\ \hline
\end{tabular}
\caption{Kinematic parameters and corresponding optimal pitch amplitudes $\theta^*$ for the baseline cases shown in Figs.\ \ref{fig:optimized_St} (variation in $St$) and \ref{fig:optimized_k} (variation in $k$).}
\label{tab:pitchAmpsBaselineCases}
\end{table}

% \todo[inline, color=orange]{Strouhal-number plot should be scalable by a factor of $h_0$. Decide whether this is worth showing.}

%\textbf{LIST OF FIGURES}
%\begin{enumerate}
%    \item Experimental setup
%    \item Standard-deviation studies (also shows PIV FOV)
%    \item Baseline case: $St$ and $k$ variations, $v$ velocities (plus one for vorticity, for %completeness)
%    \item Baseline case: pitch about LE and TE (LE better)
%    \item Baseline case: plunge with pitch, effect of phase (out of phase better)
%    \item 4-up sketch of kinematics of proposed motion
%    \item Full vorticity field to compare pitch vs. no pitch
%    \item Wake thickness plots to prove above point and demonstrate method
%    \item Acceleration line fits for left- and right-leaning plots
%    \item Optimization 1: h0 = 60 mm, f = 4 Hz (profiles, wake width, acceleration)
%    \item Optimization 2: h0 = 20 mm, f = 4 Hz (profiles, wake width, acceleration)
%    \item Optimization 3: h0 = 40 mm, f = 4 Hz (profiles, wake width, acceleration) (optional)
%    \item Optimization 4: h0 = 40 mm, f = 5 Hz (profiles, wake width, acceleration)
%    \item Sinusoidal Profiles: Strouhal-number variation
%    \item Sinusoidal Profiles: Reduced-frequency variation
%\end{enumerate}

% Note: for single-column figure, use figure* environment

\section{Conclusions}

In this work, considerations have been outlined for the generation of gusts with sinusoidal character using a single airfoil actuated in plunge and pitch. The Theodorsen theory for unsteady aerodynamics was used in order to predict the pitch amplitude necessary to produce uniform gusts at a given Strouhal number and reduced frequency. The validity of the theory was supported both by physical arguments and by experimental investigations. The quantitative predictions given by the theory were then confirmed in experiments that measured the symmetry of the vertical-velocity profiles produced in the wake of the airfoil, and the extent of the wake of the airfoil itself within the gust region. The gust signals produced were reasonably smooth, and reached significantly higher amplitudes and reduced frequencies than those attained by other gust-generation methods. The results of this work suggest that the physics of the gust-generation problem do not require high degrees of mechanical complexity to solve, but rather can be manipulated as needed with just a single airfoil and potential flow theory.

Additional steps could be taken to improve the quality of the gust flows produced in the wake of the airfoil. Employing aifoil actuators that do not protrude into the center of the test section would greatly regularize the quality of the gust in the negative portion of the profile. A thinner airfoil, built from a less flexible material, would produce an even thinner wake, and could also be actuated more reliably over a wider range of kinematics. Measurement of the forces and moments on the airfoil would also be of interest, to see more quantitatively how unsteady loads may be reduced by these kinds of kinematics. Lastly, tests at higher Reynolds numbers would allow the gust generator to be used in larger-scale experiments.

These findings have significant implications not only for the generation of well-defined gusts in aerodynamics experiments, but also for the aerodynamic behavior of tandem-wing configurations, for example in dragonfly wings, or on a larger scale, within flocks of birds or schools of fish. The control of wake structures purely by kinematics could also apply to gust-load alleviation in rotorcraft and biologically inspired unmanned aerial vehicles (UAVs). It is important to note, however, that the trends shown in this study will break down at higher Strouhal numbers and reduced frequencies, where dynamic effects become unavoidable. Nevertheless, a foray into this sector of the parameter space would allow the results of this work to be generalized to a much wider range of applications in aerodynamics, flow control, biological propulsion, and other related fields.

\begin{acknowledgements}
The authors wish to acknowledge financial support of this study by the Sino-German Center and the Deutsche Forschungsgemeinschaft through the project TR 194/55-1: ``Flow Control for Unsteady Aerodynamics of Pitching/Plunging Airfoils". In addition, Nathan Wei was supported by the German-American Fulbright Commission with a grant in the Student Category during his stay at the TU Darmstadt. The authors would also like to extend their appreciation to the workshop staff for their assistance and professional support in the preparation of the wind tunnel and test airfoil for these experiments, and to Stefan Trie{\ss}, who characterized and optimized the linear actuators.

\end{acknowledgements}

\bibliographystyle{spbasic}      % basic style, author-year citations
\nocite{}
\bibliography{main.bib}   % name your BibTeX data base

\end{document}